\documentclass[12pt]{article}
\usepackage{ccaption}
\usepackage{amssymb}
\usepackage{times}
\usepackage{titling}
\usepackage{graphicx}
\usepackage{amsmath}
\usepackage[version=3]{mhchem}
\usepackage{color}
\usepackage{bm}
\usepackage{lineno}
\usepackage{multirow}
\usepackage{float} 

\usepackage[paper=letterpaper,twoside,top=2.0cm,bottom=2cm,left=1.5cm,
            right=1.5cm,bindingoffset=0.0cm,voffset=0.0cm]{geometry}
\renewcommand{\refname}{\subsection*{References}}

\def\cbl{}

\def\bS{\bold{S}}

\def\bp{\bold{p}}
\def\bQ{\bold{Q}}
\def\bq{\bold{q}}
\def\br{\bold{r}}
\def\bR{\bold{R}}
\def\bd{\bold{d}}
\def\bU{\bold{U}}
\def\bJ{\bold{J}}

\def\ham{\mathcal{H}}

\renewcommand{\thesection}{}
\renewcommand{\thesubsection}{\arabic{subsection}}  

\title{
{\bf  Hybridized quadrupolar excitations in the spin-anisotropic frustrated magnet FeI$_{2}$}}
\author{
 	Xiaojian Bai$^{1,\star,\dagger}$, Shang-Shun Zhang$^{2}$, Zhiling Dun$^{1}$, Hao Zhang$^{2,4}$, Qing Huang$^2$, \\ Haidong Zhou$^2$,  Matthew B. Stone$^3$, Alexander I. Kolesnikov$^3$, Feng Ye$^3$,\\ Cristian D. Batista$^{2}$,  Martin Mourigal$^{1,\ddagger}$\\
 	\normalsize{$^1$School of Physics, Georgia Institute of Technology, Atlanta, GA 30332, USA } \\
 	\normalsize{$^2$Department of Physics and Astronomy, University of Tennessee, Knoxville, TN 37996, USA} \\
 	\normalsize{$^3$Neutron Scattering Division, Oak Ridge National Laboratory, Oak Ridge, TN 37831, USA} \\
	\normalsize{$^4$Materials Science and Technology Division, Oak Ridge National Laboratory, Oak Ridge, TN 37831, USA}	\\
	\normalsize{$^{\star}$\textit{Present Address: Neutron Scattering Division, Oak Ridge National Laboratory, Oak Ridge, TN 37831, USA}}\\\\
	\normalsize{$^{\dagger}$   Email: xbai33@gatech.edu}\\
	\normalsize{$^{\ddagger}$  Email: mourigal@gatech.edu}
	}
\date{April 12, 2020}
\begin{document}
\maketitle

\baselineskip24pt

{\bf Magnetic order is usually associated with well-defined magnon excitations. Exotic magnetic fluctuations with fractional, topological or multipolar character, have been proposed for radically different forms of magnetic matter  such as spin-liquids~{\cite{savary2016quantum}}. As a result, considerable efforts have searched for, and uncovered, low-spin materials with suppressed dipolar order at low temperatures~{\cite{Nakatsuji2005,Broholmeaay0668}}. Here, we report neutron-scattering experiments and quantitative theoretical modeling of an exceptional spin-1 system -- the uniaxial triangular magnet FeI$_2$~{\cite{bertrand1974susceptibilite}} -- where a \emph{bright} and \emph{dispersive} band of mixed dipolar-quadrupolar fluctuations emerges just above a dipolar ordered ground-state. This excitation arises from anisotropic exchange interactions that hybridize overlapping modes carrying fundamentally different quantum numbers. Remarkably, a generalization of spin-wave theory to local SU(3) degrees of freedom~\cite{batista2004algebraic} accounts for all details of the low-energy dynamical response of FeI$_2$ without going beyond quadratic order. Our work highlights that quantum excitations without classical counterparts can be realized even in presence of fully developed magnetic order.}

\clearpage

\baselineskip20pt

Multipolar degrees of freedom arise naturally in condensed matter systems from anisotropic distributions of charge and magnetization. In magnetic materials, crystal electric fields and spin-orbit coupling typically imprint anisotropy on the localized magnetization distribution. The interaction between the resulting anisotropic \textit{dipole} moments produces a wealth of collective lattice phenomena ranging from Ising ferromagnetism to Kitaev quantum spin-liquids~{\cite{Takagi2019}}. A comparably less explored direction is the search for \textit{multipolar} phases that lack magnetic (dipolar) order {\cite{suzuki2018first}. Multipolar order, often referred as ``hidden order", is evasive to most conventional probes of materials. Thus, experimental realizations in the solid-state are rare and limited to a small number of $f$-electron systems~{\cite{kuramoto2009multipole,santini2009multipolar}}. An alternative route to study multipolar physics is to consider systems for which magnetic order is present, but multipolar fluctuations coexist with dipolar excitations at low energies. In that case, it becomes convenient to represent the fundamental degrees of freedom using a SU(N) spin~\cite{batista2004algebraic}. This representation treats all multipolar components on equal footing and predicts radically new types of quantum excitations~\cite{matsumoto2007longitudinal,romhanyi2012multiboson}. Quantitative tests of that approach are generally lacking, however, because multipolar fluctuations tend to be silent in spectroscopy experiments.

Here, we demonstrate that the magnetically ordered triangular-lattice compound FeI$_2$ -- a system known and studied since the 1970's \cite{fert1978excitation, petitgrand1980magnetic} -- is an exceptional platform to observe and model multipolar fluctuations. Although the magnetism of FeI$_2$\ results from complex exchange interactions, which we elucidate in detail further below, the essence of its multipolar physics is captured by a simple $S\!=\!1$ chain with ferromagnetic Heisenberg exchange $J$ and strong easy-axis single-ion anisotropy $D\!\gg\!|J|$ [Fig.~\ref{fig:1}]. In that model, low-energy excitations comprise two types of fluctuations carrying fundamentally different quantum numbers [Fig.~\ref{fig:1}{\bf a}]: conventional magnons corresponding to the propagation of a single $|S^{z}\!=\!0\rangle\equiv|0\rangle$ defect in a ferromagnetic background of $|+\!1\rangle$ sites, and single-ion bound-states (SIBS)~\cite{silberglitt1970effect,oguchi1971theory} for which a spin on a given site is flipped from $|+\!1\rangle$ to $|-\!1\rangle$, with an energy that is independent of $D$. Dipolar matrix elements vanish for SIBS excitations, because the operator ${S}^{-}\!=\!{S}^x\!-\!i{S}^y$ must act on the same site twice. As a consequence, the latter cannot be detected using experimental probes bound by the dipole selection rule. The multipolar nature of SIBS excitations is readily understood using a SU(N) spin representation, where local $S\!=\!1$ degrees of freedom are represented by eight linearly-independent SU(3) generators. The first three are the usual dipolar operators $S^\mu$, while the remaining five are the quadrupolar operators ${ Q}^{\mu \nu}=({S}^{\mu} {S}^{\nu} + {S}^{\nu} {S}^{\mu})/2 -2/3 {\delta^{\mu\nu}}$. In this description, SIBS excitations are local quadrupolar fluctuations generated by ${Q}^{xx}-{Q}^{yy}-2i{Q}^{xy}\!\equiv\!({S}^{-})^{2}$, and the single ion term $-D({S}_{i}^{z})^2$ coincides with ${Q}^{zz}$ up to a constant.

\begin{figure}[h!] 
	\begin{center}
		\includegraphics[width=0.8\textwidth]{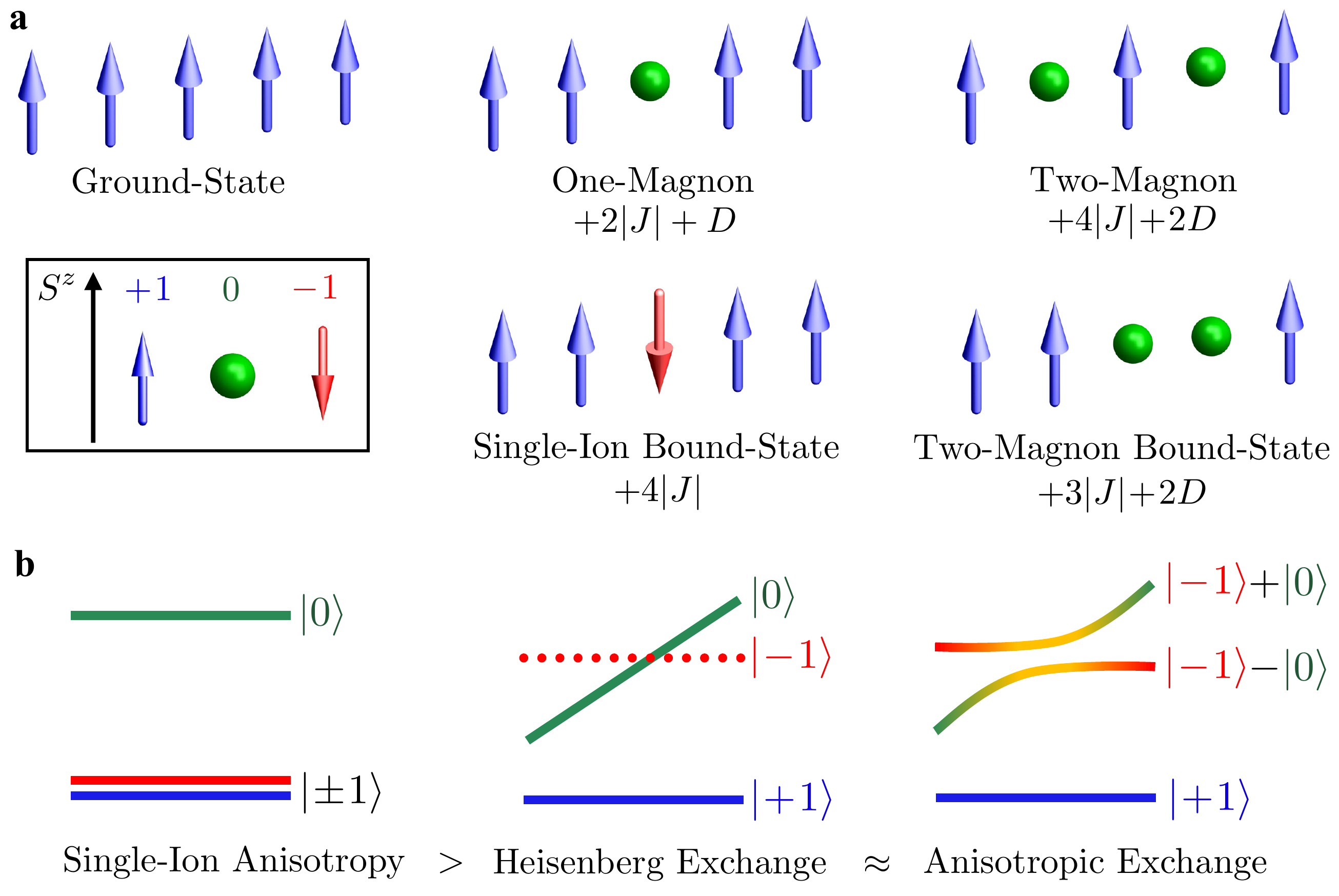}
	\end{center}
	\caption{{\bf Elementary magnetic excitations of a ferromagnetic easy-axis spin-1 chain and their hybridization through anisotropic exchange interactions.} {\bf a.} Sketches of the ground-state, single-excitation, and double-excitations of a model spin-1 Hamiltonian, $\mathcal{H} = J\sum_{\langle i,j\rangle}\bS_{i}\!\cdot\!\bS_{j}-D\sum_i (S_{i}^{z})^2$, where $J\!<\!0$ is a ferromagnetic nearest-neighbor exchange interaction and $D\!>\!0$ represents an easy-axis single-ion anisotropy. For an individual spin ($S\!=\!1$), three states are possible corresponding to projections $S^{z}\!=\!+1$, $0$, or $-1$, which are represented by an up-arrow, a dot and a down-arrow, respectively. In a local picture, elementary excitations of the ferromagnetic ground-state comprise the creation of a single magnon ($|S^z\!=\!+1\rangle\!\equiv\!|+\!1\rangle\!\to\!|0\rangle$, $|\Delta S^z|\!=\!1$) with energy $2|J|+D$, and three types of doubly-excited states. A two-magnon excitation generally involves two $|0\rangle$ states on non-adjacent sites and costs an energy $4|J|+2D$. The energy of such an excitation can be lowered by the formation of one of two possible bound-states: a two-magnon bound-state (TMBS) for which two $|0\rangle$ states reside on neighboring sites and experience an effective off-site attraction $-|J|$ (total energy $3|J|+2D$); and a single-ion bound-state (SIBS) for which the same site is excited twice ($|+\!1\rangle\!\to\!|-\!1\rangle$, $|\Delta S^z|\!=\!2$) with an effective on-site attraction $-2D$ (total energy $+4|J|$). For sufficiently strong anisotropy, $D\!\gg\!|J|$, the SIBS and the one-magnon are the dominant low-energy excitations. {\bf b.} Effect of exchange interactions on low-energy excitations. Starting from a doubly-degenerate single-ion ground-state, Heisenberg (diagonal) exchange interactions lift the ground-state degeneracy through magnetic ordering. The one-magnon excitation acquires a momentum-energy dispersion, while the SIBS remains localized and invisible to conventional spectroscopic probes due to a vanishing dipolar matrix element. Anisotropic (symmetric off-diagonal) exchange interactions hybridize the single-magnon and SIBS excitations at their crossing points, making the SIBS visible and dispersive.}
	\label{fig:1}
\end{figure}

Surprisingly, quadrupolar SIBS fluctuations in FeI$_2$ are easily detected by magneto-infrared spectroscopy through a doubled $g$-factor compared to magnon modes \cite{fert1978excitation, petitgrand1980magnetic}, inelastic neutron scattering \cite{petitgrand1979neutron}, and electron-spin resonance \cite{katsumata2000single}. To understand the detailed mechanism enabling the observation of such quadrupolar fluctuations with dipolar spectroscopic probes, we combine state-of-art inelastic neutron scattering experiments on large single-crystals of FeI$_2$ with a quantitative SU(3) generalized spin-wave theory (GSWT) model, see Methods and Supplementary Information. Our results uncover a genuine quantum-mechanical effect active in spite of a magnetically ordered ground-state and negligible longitudinal fluctuations: single-ion anisotropy, Heisenberg exchange and anisotropic interactions cooperate to produce an accidental overlap and subsequent hybridization between dipolar and quadrupolar fluctuations. This opens up an energy gap and transforms the otherwise dark and flat SIBS excitation into a bright and dispersive mode [Fig.~\ref{fig:1}{\bf b}]. 

This phenomenon occurs in FeI$_2$ due to spatially-complex exchange pathways between magnetic Fe$^{2+}$ ions residing on perfect triangular-lattice layers and spin-orbit effects within the weakly-distorted trigonal environment of I$^-$ ligands [Fig.~\ref{fig:2}{\bf a}]. At the single-ion level, a well-isolated triplet is stabilized below 10~meV [Fig.~\ref{fig:2}{\bf b}] and maps onto an effective $S\!=\!1$ model with an easy-axis single-ion anisotropy~\cite{balucani1985hybrid} $D\approx1.9$ to $2.2$~meV~\cite{bertrand1974susceptibilite,fujita1966mossbauer}. The magnetic exchange interactions are an order of magnitude smaller~\cite{petitgrand1980magnetic} and stabilize a collinear $c$-axis magnetic order~\cite{gelard1974magnetic, wiedenmann1988neutron} below $T_N\!=\!9.5$~K in zero magnetic field, through a first order transition with no apparent lattice distortion~\cite{gelard1974magnetic}. The magnetic structure features an up-up-down-down stripe configuration in the $ab$-plane, which shifts by one unit along the $a$-axis between subsequent triangular layers [Fig.~\ref{fig:2}{\bf c}]. With a propagation vector ${\bf k}_m\!=\!(1/4,0,1/4)$, this yields three magnetic domains in the material, related by 120$^\circ$ rotations, which exist with different fractions in our large single-crystals [Fig.~\ref{fig:2}{\bf d} and Supplementary Section 3].

To quantitatively determine the exchange interactions in FeI$_2$, we construct a minimal model which realizes the observed magnetic structure at low temperature [Fig.~\ref{fig:2}{\bf a}]. Due to the dominant easy-axis anisotropy, it suffices to consider an Ising Hamiltonian, $\mathcal{H}=\sum_{(i,j)} J_{ij} {S}_{i}^{z}{S}_{j}^{z}$, and find constraints on $J_{ij}$ that stabilize the observed magnetic order. Within the triangular plane, the up-up-down-down stripe structure requires competing exchange interactions between first, second and third neighbors \cite{tanaka1975ground}, with $J_1\!<0$ ferromagnetic, $J_2\!>0$ antiferromagnetic, and $J_3$ such that $J_1-2J_3<0$ and $J_1+2J_2+2J_3>0$. By enumerating all possible 3D stacking sequences with a $c$-axis periodicity at most four times that of the crystal [Supplementary Section 4], we obtain the required magnetic structure if $J^{'}_{2a}$ is antiferromagnetic and $-J^{'}_{2a}<J^{'}_{0}+2J^{'}_{1} <3J^{'}_{2a}$. Distinguishing between equal-length but symmetry-inequivalent interactions of $J^{'}_{2a}$ and $J^{'}_{2b}$ is crucial in removing otherwise degenerate solutions. Moreover, $J^{'}_{2a}$ corresponds to an almost 180$^\circ$ Fe-I-I-Fe bridge, while $J^{'}_{2b}$ cannot not be reasonably associated with a super-exchange pathway; we neglect it from here on. In this Ising approach, flipping a spin costs an energy $4(-J_1+J_2+J_3+2J_{2a}^{'})$, which is contrained by the experimental SIBS energy of ${\approx}\ 2.8$~meV~{\cite{fert1978excitation}}.

\begin{figure}[th!] 
	\begin{center}
		\includegraphics[width=0.9\textwidth]{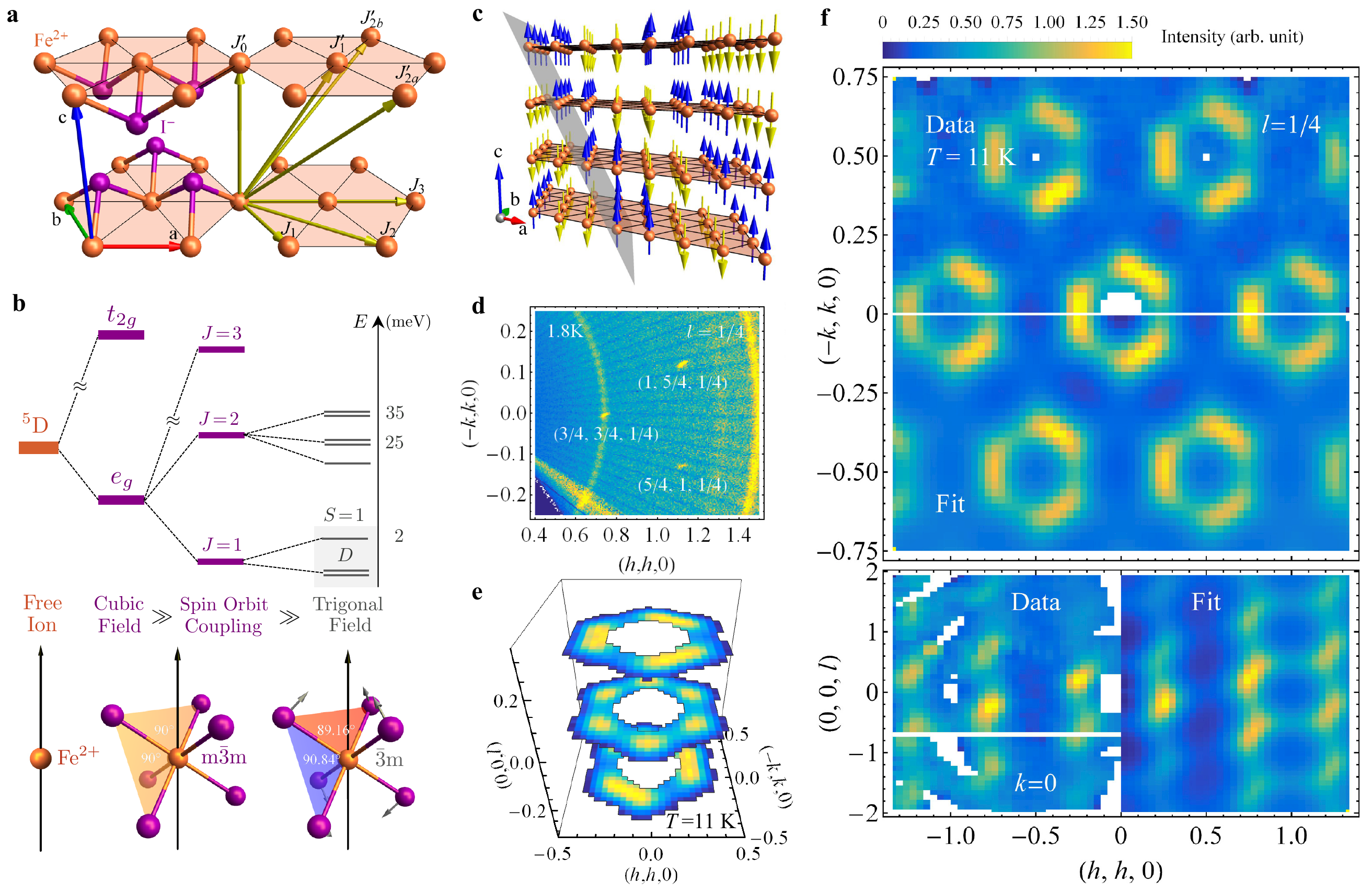}
	\end{center}
	\caption{
	{\bf Microscopic origin of the magnetic properties of FeI$_2$.} 
	{\bf a.} Trigonal crystal structure of FeI$_2$ (space-group $P\bar{3}m1$, $a\!=\!4.05$~\AA~and $c\!=\!6.75$~\AA~ at $T\!=\!300$~K), showing triangular layers of Fe$^{2+}$ ions and resulting symmetry-inequivalent magnetic exchange pathways, mediated by I$^-$ ligands, up to third nearest neighbors in-plane $(J_1,J_2,J_3)$ and out-of-plane $(J^{'}_{0},J^{'}_{1},J^{'}_{2a},J^{'}_{2b})$. {\bf b.} Local coordination environment of Fe$^{2+}$ ions and hierarchy of single-ion energy-scales. Starting from a $L\!=\!2, S\!=\!2$ free-ion state~\cite{lockwood1994raman}, the dominant cubic crystal field ($\approx1$~eV) splits the fivefold degenerate $3d$-orbitals into a ground-state $t_{2g}$ triplet ($\ell_{\rm eff}\!=\!1$) and excited $e_g$ doublet~\cite{fujita1966mossbauer,balucani1985hybrid}. The sub-leading spin-orbital coupling (${\approx} 10$~meV) lifts the resulting $(2S+1)\!\times\!(2\ell_{\rm eff}+1)\!=\! 15$-fold degeneracy into three multiplets with total angular momentum $J = 1, 2,$ and $3$. The inter-multiplet transition $J\!=\!1\rightarrow2$ is observed between 25 and 35~meV~\cite{lockwood1994raman}. At low energies, the weak trigonal distortion from $m\bar{3}m$ to $\bar{3}m$ ( ${\approx} 1$~meV) maps onto an easy-axis anisotropy for the effective $S=1$ ground-state. {\bf c.} Magnetic structure of FeI$_2$, showing ferromagnetic planes (gray) arranged in a up-up-down-down (blue-blue-yellow-yellow) sequence. {\bf d.} Elastic neutron-scattering intensity collected at $T\!=\!1.8$~K using the CORELLI instrument, showing magnetic Bragg peaks in the ${\bf Q}\!=\!(h,k,1/4)$ plane from three equivalent ${\bf k}$-magnetic domains related by $120^{\circ}$ rotations. {\bf e.} Diffuse neutron-scattering intensity collected at $T\!=\!11$~K on the SEQUOIA instrument, obtained by integrating energy transfer from $0$ to $6$~meV, showing cuts through the three-dimensional distribution of scattering intensity consistent with the $\bar{3}m$ Laue symmetry. {\bf f.} Extended cuts through the diffuse-scattering data in the $(h,k,1/4)$ (top) and $(h,h,l)$ planes (bottom) and comparison to SCGA calculations with fitted parameters, see text and Methods.}
	\label{fig:2}
\end{figure}

To gain further insight, we extend the Heisenberg model of Fig.~\ref{fig:1} to the realistic exchange interactions of Fig.~\ref{fig:2}{\bf a}, and use the self-consistent Gaussian approximation (SCGA) \cite{conlon2010absent} to model the highly-structured diffuse scattering data collected in the paramagnetic phase at $T\!=\!11$~K [Fig.~\ref{fig:2}{\bf e}]. This approach has proven to be very successful in extracting quantitative microscopic interactions for frustrated magnets \cite{plumb2019continuum,bai2019magnetic}. Here, we perform a global fit to the paramagnetic data constraining the $J_{ij}$ parameters to reproduce the SIBS energy. We obtain a good agreement between SCGA calculations and the entire momentum-dependence of our data [Fig.~\ref{fig:2}{\bf e}], see Methods for details and fit results. The fitted exchange parameters necessarily satisfy all the stability conditions of the ordered structure. Notably, the magnitude of $J_1\!=\!-0.24$~meV is comparable to $J_2\!=\!0.11$~meV and $J_3\!=\!0.21$~meV, which can be attributed to a cancellation of ferromagnetic direct-exchange and antiferromagnetic super-exchange between nearest-neighbors \cite{Wu2012}. 

We then proceed to calculate the energy-resolved response of this model using linear spin-wave theory (LSWT) and compare it with our high-resolution inelastic neutron-scattering data [Fig.~\ref{fig:3}{\bf a}]. Despite the overall resemblance, this approach fails in several important aspects. First, while the data contains two separate bright excitation bands, corresponding to SIBS and one-magnon excitations, the former is completely missed by LSWT, which ignores quadrupolar degrees of freedom from the start. In conventional spin-wave theory, bound-states can be obtained by introducing magnon-magnon interactions but require summation of ladder diagrams up to infinite order in $1/S$ expansion, or equivalently solving a set of complicated integral equations~{\cite{oguchi1971theory}}. Instead, we employ the generalized spin-wave theory (GSWT) and represent the local states of SU(3) spins using Schwinger bosons \cite{muniz2014generalized}, $|m\rangle_{i} = b^{\dagger}_{i,m}|\text{vac}\rangle_{i}$ where $m=+1,0,-1$ refers to the quantized angular momentum along the $c$-axis, $i$ labels the lattice sites, and we enforce the constraint $\sum_{m}b^{\dagger}_{i,m}b^{}_{i,m}=1$. For the magnetically ordered ground-state of FeI$_2$, $b^{\dagger}_{i,+1}$ bosons are condensed, $b^{}_{i,0}$ creates a single magnon, and $b^{}_{i,-1}$ creates a SIBS. In the GSWT approach [Supplementary Section 6], diagonalizing a quadratic Hamiltonian is sufficient to capture both dipolar and quadrupolar excitations, but they remain decoupled for purely diagonal (Heisenberg) exchange interactions because they carry different quantum numbers. This implies that the SIBS excitation is completely flat, localized, and invisible in neutron scattering experiments. Thus, a second hitherto unexplained aspect of our data is the dispersive nature of the SIBS excitation wherever it approaches the one-magnon band [Fig.~\ref{fig:3}{\bf a}]. Given the overlap between the calculated SIBS energy and LSWT magnon dispersion, our data suggests that a strong hybridization occurs between dipolar and quadrupolar fluctuations in FeI$_2$ [Fig.~\ref{fig:1}{\bf b}]. 

\begin{figure}[th!] 
	\begin{center}
		\includegraphics[width=0.80\textwidth]{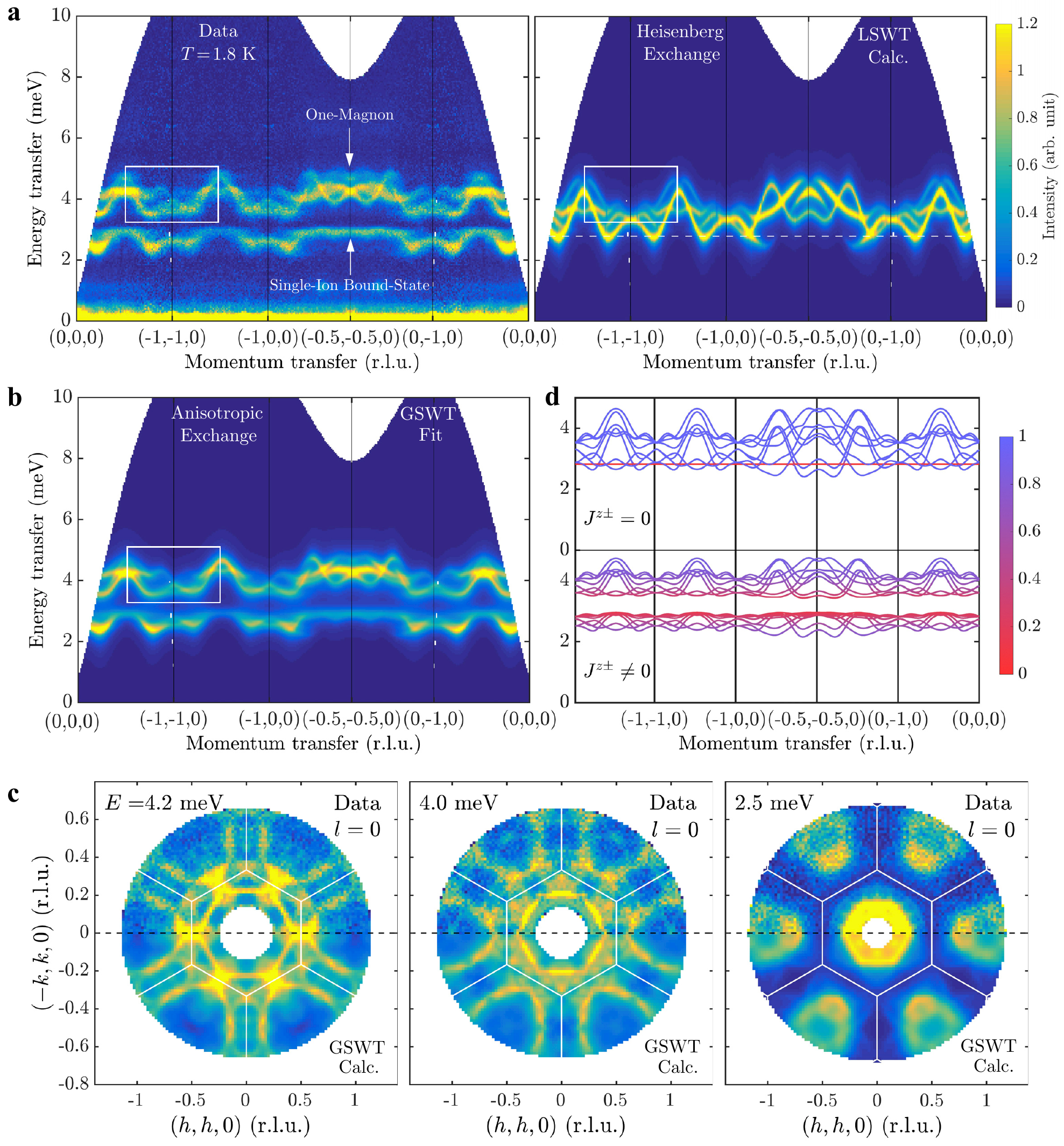}
	\end{center}
	\caption{\label{fig4}{\bf Low-energy magnetic excitations of FeI$_2$ and matching anisotropic exchange model for SU(3) spins.} {\bf a.} (Left) Energy-resolved neutron-scattering intensity collected at $T\!=\!1.8$~K using the SEQUOIA spectrometer, showed along a high-symmetry path in the triangular Brillouin zone after data symmetrization and integration over $l\leq0.1$ r.l.u. (Right) Corresponding linear spin-wave theory calculation for our Heisenberg model with exchange parameters fitted from diffuse-scattering data of Fig.~\ref{fig:2}{\bf f}. The white dashed line indicates the SIBS energy\cite{fert1978excitation}. White boxes indicate symmetry-equivalent positions in reciprocal space with large intensity disparities that are not explained by a Heisenberg model with the usual Fe$^{2+}$ form factor. {\bf b.} Generalized spin-wave theory fit to the data using our anisotropic model. For a comparison to cuts in the out-of-plane direction, see Supplementary Section 7. {\bf c.} Momentum-transfer dependence of magnetic excitations at selected energies (top panels) and GSWT calculations (bottom panels) using the best fitting parameters of the anisotropic model.  {\bf d.} Hybridization effect from anisotropic exchange interaction $J^{z\pm}$ represented through the relative weight of $b^{\dagger}_{i,-1}$ (SIBS, red) and $b_{i 0}$ (one-magnon, blue) in a given excitation eigenvector.}
	\label{fig:3}
\end{figure}

A hint at the hybridization mechanism comes from our observation of an anomalous distribution of scattering intensity across reciprocal space. Starting from the $\Gamma_2\!\equiv\!(-1,-1,0)$ point of the hexagonal Brillouin zone, the experimental intensity for momentum-transfer along the path $\Gamma_2\!\rightarrow \Gamma_1 \equiv (0,0,0)$ differs substantially from the symmetry-related path $\Gamma_2\!\rightarrow\!\Gamma_2^\prime \equiv (-1,0,0)$ [white boxes in Fig.~\ref{fig:3}{\bf a}]. This phenomenon is totally absent in our LSWT calculations and results from the combined effect of anisotropic exchange interactions and neutron dipole factor~{\cite{paddison2020scattering}}. A symmetry analysis of spin-space anisotropy for nearest-neighbor bonds yields four independent parameters,  such that the corresponding exchange Hamiltonian can be written
\begin{align}\label{eq:NNHam}
\mathcal{H}_{\text{n.n.}}  =  \sum_{\left\langle i,j\right\rangle }&\big\{J_{1}^{zz}{S}_{i}^{z}{S}_{j}^{z}+J_{1}^{\pm}\left({S}_{i}^{+}{S}_{j}^{-}+{S}_{i}^{-}{S}_{j}^{+}\right)
	 + J_{1}^{\pm\pm}\left(\gamma_{ij}{S}_{i}^{+}{S}_{j}^{+}+\gamma_{ij}^{*}{S}_{i}^{-}{S}_{j}^{-}\right) \nonumber\\
	& -\dfrac{iJ^{z\pm}_{1}}{2}\left[ (\gamma^{*}_{ij}{S}^{+}_{i}-\gamma_{ij}{S}^{-}_{i}){S}^{z}_{j}+ {S}^{z}_{i}(\gamma^{*}_{ij}{S}^{+}_{j}-\gamma_{ij}{S}^{-}_{j})\right]\big\}\,, \nonumber
\end{align}
where $\gamma_{ij}$ are bond-dependent phase factors [Supplementary Section 8]. This contribution to the Hamiltonian of FeI$_2$ has the same form as proposed for the rare-earth triangular-lattice compound YbMgGaO$_4$~{\cite{li2015rare,maksimov2019anisotropic}} for which an intensity modulation in the diffuse scattering data was also reported \cite{paddison2017continuous}. To progress, we introduce a \textit{minimal} 15-parameters model that includes all four anisotropic exchange parameters on nearest-neighbor bonds, diagonal XXZ anisotropy for the five non-negligible further-neighbor interactions, and single-ion anisotropy. We calculate the scattering intensity for this model using GSWT, perform a pixel-to-pixel fit to data along high-symmetry paths in the $(h,k,0)$-plane and for selected cuts in the $l$-direction, and use the rest of the data as a check. The fit quality is excellent [Fig.~\ref{fig:3}{\bf b}], including anomalous intensities, and the obtained parameters allow to compute constant energy slices that are in remarkable agreement with the data [Fig.~\ref{fig:3}{\bf c}], see Methods for details. 

Our analysis uncovers that the symmetric off-diagonal exchange interaction $J_{1}^{z\pm}\!=\!-0.260(1)$~meV is larger than any transverse exchange $J_{i}^{\pm}$ in the system and responsible for hybridizing the overlapping one-magnon and SIBS bands, and opening an energy gap between them. At quadratic order in Schwinger boson representation, ${S}_{i}^{z}{S}_{j}^{+}$ maps onto $(b^{\dagger}_{i,-1}b_{i, 0}+b^{\dagger}_{j,-1}b_{j, 0})$, which introduces an on-site coupling transforming a single magnon into a SIBS. As a result, these excitations acquire a mixed dipolar-quadupolar character throughout the Brillouin zone [Fig.~\ref{fig:3}{\bf d}], which explains the bright and dispersive parts of the SIBS branch as inherited from the one-magnon mode. Our calculations also suggest that $J_{1}^{z\pm}$ rotates the orientation of dipoles in the collinear magnetic structure $\approx 10^\circ$ from the $c$-axis [Supplementary Section 7]. For the parameters of FeI$_2$, the SU(3) ground-state wave-function remains in close proximity to a SU(2) coherent state due to the dominant single-ion anisotropy. As a result, local expectation values have an entirely dipolar character with negligible quadrupolar contributions [Supplementary Section 6] and the hybridized excitations are almost entirely transverse to the direction of the dipolar moments. In summary, our work uncovers that anisotropic exchange interactions in FeI$_2$ hybridize dipolar and quadrupolar fluctuations via a non-perturbative quantum-mechanical phenomenon that we explain quantitatively for the first time. Furthermore, we demonstrate that generalizing spin-wave theory to SU(3) degrees of freedom allows to quantitatively match high-resolution neutron scattering data at quadratic order, which would otherwise require the treatment of magnon-magnon interactions to infinite order in a $1/S$-expansion.

Amongst transition-metal systems with partially-quenched orbital degrees of freedom, excitations with multipolar characters have only been conclusively detected in a handful of compounds such as Ba$_2$CoGe$_2$O$_7$ \cite{penc2012spin}, Sr$_2$CoGe$_2$O$_7$ \cite{akaki2017direct}, NiCl$_2$-4SC(NH$_2$)$_2$ \cite{zvyagin2008observation} and CsFeCl$_3$ \cite{yoshizawa1980neutron,hayashida2019novel}. Yet, FeI$_2$ stands out as the only realization with strong easy-axis anisotropy and, consequently, quadrupolar fluctuations appearing almost purely in the transverse channel. While it is fortuitous that dipolar and quadrupolar excitations overlap in FeI$_2$, it appears promising to search for multipolar excitations in other large-spin systems using tuning parameters such as magnetic field or pressure. In light of our results, the very nature of magnetic excitations in FeI$_2$ below the saturation magnetic field of $\approx$12.5T  \cite{fert1973phase} calls for further investigation, given the rich sequence of magnetic structures observed in neutron diffraction~\cite{wiedenmann1988neutron,wiedenmann1989magnetic}. Overall, perhaps the most important message of our work is to highlight that anisotropic large-spin systems such as FeI$_2$ or the recently studied $\alpha$-NaMnO$_2$~{\cite{dally2020high}}, are not to be shunned away from detailed studies for {\cbl supposedly lacking of quantum effect, and novel quantum excitations without classical counterparts can be realized in a fully ordered magnet.}

\clearpage
\subsection*{References}
\begingroup
\renewcommand{\section}[2]{}%

\endgroup

\subsection*{Methods}
\baselineskip20pt

\noindent\textit{Sample preparation.} 
Small single-crystal samples of FeI$_2$ were grown in evacuated quartz tubes from pure elements using the chemical vapor transport technique with the hot end at 570$^\circ$C and the cold end at room temperature \cite{method-coleman1993optimization}. As-grown FeI$_2$ crystals appear as thin black flakes, very easy to bend and cut. Due to their highly hygroscopic nature, all samples were handled in a glovebox. Small crystals were collected and sealed in quartz tubes under vacuum. Large single crystals up to 3.0 grams were grown by slowly passing the resulting tube through a floating zone furnace at high temperature. In addition to the references in the main text, the magnetic properties of FeI$_2$ are discussed in Refs.~\cite{method-petitgrand1976far,method-friedt1976electronic,method-trooster1978spin,method-katsumata2000observation}.\\ 

\noindent\textit{X-ray diffraction measurements and refinements.} 
Room-temperature powder X-ray diffraction (PXRD) were carried out on crushed single-crystal samples using a PANAnalytical Empyrean Cu-$K\alpha$ diffractometer. Due to their highly hydroscopic nature, FeI$_2$ crystals degrade within a few seconds when exposed to air. Samples were loaded in an air-tight domed holder in the glovebox to keep them from degrading during PXRD measurement. Rietveld refinement was carried out using the FULLPROF program \cite{method-rodriguez1993recent} with fits to the data and refined values of structural parameters in Supplementary Information, Section 1.\\ 

\noindent\textit{Neutron scattering measurements.}
Elastic neutron-scattering experiments were performed using the CORELLI spectrometer at the Spallation Neutron Source (SNS), Oak Ridge National Laboratory (ORNL), USA. A thin-flake sample of mass $m\!=\!0.4$~g was sealed in aluminum foil an mounted on an aluminum holder. The sample was cooled in a liquid helium cryostat reaching a base temperature of $T\!=\!1.8$~K. The sample was rotated in steps of $3^{\circ}$ with a range of $82^{\circ}$. The white-beam Laue method provided access to large volumes of reciprocal space while the cross-correlation method was used to reconstruct the elastic signal~{\cite{method-ye2018implementation}}. Inelastic neutron-scattering experiments were performed on the SEQUOIA time-of-flight spectrometer at SNS, ORNL, USA~\cite{method-granroth2010sequoia,method-stone2014comparison}. The sample was a $m\!=\!2.5$~g slab crystal sealed in aluminum foil, mounted on an aluminum holder, and aligned with the $ab$-plane horizontal using the CG-1B alignment station at the High Flux Isotope Reactor (HFIR), ORNL, USA. The sample holder was attached to a sample stick inserted in a liquid helium cryostat reaching a base temperature of $T\!=\!1.8$~K. Measurements at $T\!=\!1.8$~K (respectively $11$~K) were performed by rotating the sample in steps of $0.5^\circ$ (respectively $1^{\circ}$) using $E_i\!=\!12$~meV (respectively $65$~meV) and choppers in high-resolution mode yielding a FWHM elastic energy resolution of $0.27$~meV (respectively $1.76$~meV). \\

\noindent\textit{Data reduction and analysis.} Initial data reduction was performed in MANTID~\cite{method-arnold2014mantid} for both SEQUOIA and CORELLI datasets. Throughout the manuscript, the scattering intensity is measured as a function of energy transfer $E$ and momentum-transfer ${\bf Q} = h {\bf a^\ast} +   k {\bf b^\ast} + l {\bf c^\ast} \equiv (h,k,l)$ where ${\bf a}^\ast$,  ${\bf b}^\ast$ and ${\bf c}^\ast$ are the primitive vectors of the reciprocal space. The conventions is such that ${\bf a}^\ast$ and ${\bf b}^\ast$ makes an $120^\circ$ angle [Supplementary Section 2 and 7]. Subsequent analysis of the SEQUOIA data was performed in HORACE \cite{method-ewings2016horace} on a dedicated node within Georgia Tech's Partnership for Advanced Computing infrastructure. The data was symmetrized using the $\bar{3}m$ Laue symmetry [Supplementary Section 2], a procedure which averages over all magnetic domains in the ordered phase. \\

\noindent\textit{SCGA modeling and fit results.} We modeled the diffuse scattering intensity using the Self-Consistent Gaussian Approximation (SCGA) for the Heisenberg model of Fig.~\ref{fig:1} extended to the exchange interactions of Fig.~\ref{fig:2}{\bf a}, $\mathcal{H} = \sum_{(i,j)}J_{ij}\, \bS_{i} \cdot \bS_{j}-D\sum_i (S_{i}^{z})^2$. The fit to the diffuse scattering data in Fig.~\ref{fig:2}{\bf f} yields $J_1 = -0.24$~meV, $J_2 = 0.11$~meV, $J_3 = 0.21$~meV, $J_0^{'} = -0.04$~meV, $J_1^{'} = 0.05$~meV, $J_{2a}^{'} = 0.07$~meV, and $D=2.17$~meV. We verified that for these fit results, the SCGA matches with more accurate classical Monte-Carlo results for our experimental temperature of $T=11$~K [Supplementary Section 5].\\
 
\noindent\textit{Generalized Spin Wave Theory and Fitting.} We modeled the inelastic neutron scattering response by calculating the dynamical structure factor using a SU(3) Schwinger boson representation of the spin operators [Supplementary Section 6] and the usual form-factor of Fe$^{2+}$. Using a 15-parameters model, we obtain an excellent fit [Supplementary Section 7] to all the data [Fig.~\ref{fig:2}{\bf f}] for the parameters $J^{\pm}_1=-0.119(1)$~meV, $J^{\pm\pm}_1=-0.087(4)$~meV, $J^{z\pm}_1=-0.260(1)$~meV, $J^{\pm}_2=+0.012(1)$~meV, $J^{\pm}_3=+0.084(1)$~meV, $J^{'\pm}_0=+0.014(0)$~meV, $J^{'\pm}_1=+0.007(0)$~meV, $J^{'\pm}_{2a}=+0.031(0)$~meV and $D=+2.21(2)$~meV. The $J^{zz}$ parameters are determined with higher uncertainty: $J^{zz}_1=-0.212(77)$~meV, $J^{zz}_2=+0.062(86)$~meV, $J^{zz}_3=+0.407(85)$~meV, $J^{'zz}_0=0$~meV (fixed), $J^{'zz}_1=0$~meV (fixed) and $J^{'zz}_{2a}=+0.012(34)$~meV, because the spectrum is only sensitive to the combination $E_{\text{SIBS}}=4(-J^{zz}_1+J^{zz}_2+J^{zz}_3+2J_{2a}^{'zz})$, the energy of the single-ion bound state.\\

\subsection*{Method References}
{\def\section*#1{}
\renewcommand\refname{}

}


\subsection*{Acknowledgements}
\baselineskip20pt
We thank Collin Broholm, Itamar Kimchi, Stephen Nagler, Oleg Starykh, and Alan Tennant for valuable discussions. The work of X.B., Z.D., and M.M. at Georgia Tech was supported by the U.S. Department of Energy, Office of Science, Basic Energy Sciences, Materials Sciences and Engineering Division under award DE-SC-0018660. The work of S.-S.Z. and C.D.B. at the University of Tennessee was supported by the Lincoln Chair of Excellence in Physics and the work of H.Z. was supported by the Department of Energy. The work of Q.H and H.D.Z. at the University of Tennessee was supported by the U.S. Department of Energy, Office of Science, Basic Energy Sciences, Materials Sciences and Engineering Division under award DE-SC-0020254. The research at Oak Ridge National Laboratory's Spallation Neutron Source and High Flux Isotope Reactor was sponsored by the U.S. Department of Energy, Office of Basic Energy Sciences, Scientific User Facilities Division.

\subsection*{Author Contributions}
\baselineskip20pt
X.B. and M.M. conceived the project, which was supervised by C.D.B. and M.M.. Z.D. and X.B. grew the samples with the assistance of Q.H. and H.D.Z. using the floating-zone furnace. Z.D. aligned the sample for measurements. X.B., Z.D., M.B.S., A.I.K., F.Y. and M.M. performed the neutron-scattering measurements. X.B. analyzed the data and performed fits. S.-S.Z., H.Z. and C.D.B. carried out the GSWT calculations and assisted  with the theoretical interpretation. S.-S.Z. made the GSWT code used to fit the inelastic spectra. X.B. and M.M. wrote the manuscript with input from all authors.



       \renewcommand\refname{References}
       \renewcommand{\thesection}{S.\arabic{section}}
       \renewcommand{\thesubsection}{\thesection.\arabic{subsection}}
        \setcounter{equation}{0}
        \makeatletter 
        \renewcommand{\theequation}{S\arabic{equation}}
        \makeatother
        \setcounter{figure}{0}
        \makeatletter
        \makeatletter \renewcommand{\fnum@figure}
        {\figurename~S\thefigure}
        \makeatother
        \setcounter{table}{0}
        \makeatletter
        \makeatletter \renewcommand{\fnum@table}
        {\tablename~S\thetable}
        \makeatother

\title{\vspace{-1cm}
{\bf  Hybridized quadrupolar excitations in the spin-anisotropic frustrated magnet FeI$_{2}$}\\
{\it Supplementary Information}}
\author{
 	Xiaojian Bai$^{1,\star,\dagger}$, Shang-Shun Zhang$^{2}$, Zhiling Dun$^{1}$, Hao Zhang$^{2,4}$, Qing Huang$^2$, \\ Haidong Zhou$^2$,  Matthew B. Stone$^3$, Alexander I. Kolesnikov$^3$, Feng Ye$^3$,\\ Cristian D. Batista$^{2}$,  Martin Mourigal$^{1,\ddagger}$\\
 	\normalsize{$^1$School of Physics, Georgia Institute of Technology, Atlanta, GA 30332, USA } \\
 	\normalsize{$^2$Department of Physics and Astronomy, University of Tennessee, Knoxville, TN 37996, USA} \\
 	\normalsize{$^3$Neutron Scattering Division, Oak Ridge National Laboratory, Oak Ridge, TN 37831, USA} \\
	\normalsize{$^4$Materials Science and Technology Division, Oak Ridge National Laboratory, Oak Ridge, TN 37831, USA}	\\
	\normalsize{$^{\star}$\textit{Present Address: Neutron Scattering Division, Oak Ridge National Laboratory, Oak Ridge, TN 37831, USA}}\\\\
	\normalsize{$^{\dagger}$   Email: xbai33@gatech.edu}\\
	\normalsize{$^{\ddagger}$  Email: mourigal@gatech.edu}
	}
\maketitle	
\tableofcontents
\clearpage
\section{Powder X-ray diffraction measurements}
\begin{figure}[th!]
\centering
  \includegraphics[width=0.8\textwidth]{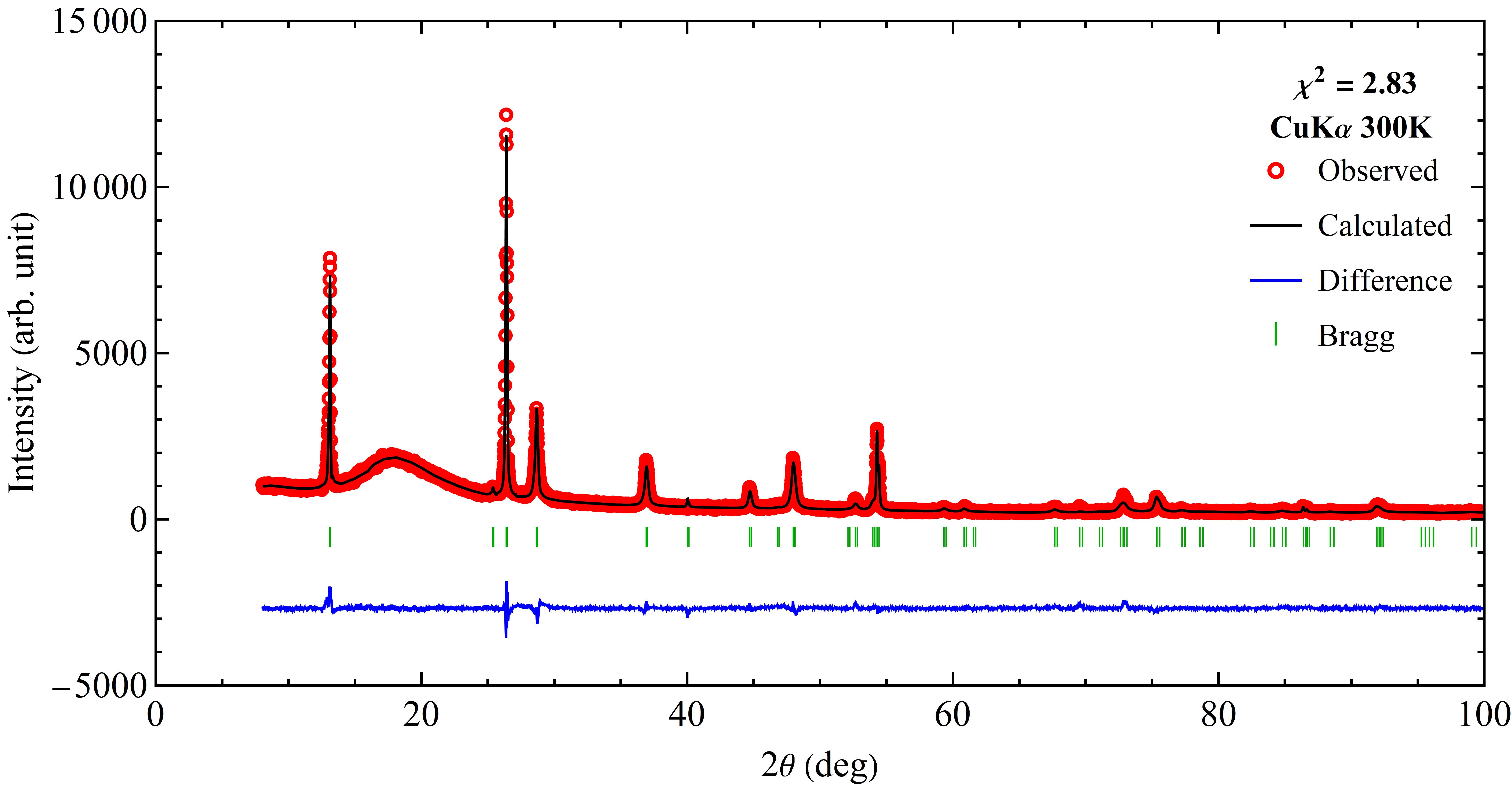}
  \caption{Room-temperature X-ray diffraction patterns of crushed single-crystals measured in a domed sample holder spinning at $16$~RPM. The broad peak at $2\theta\sim 20^{\circ}$ is the background from the polycarbonate dome which keeps the sample from degrading during the measurement. Strong preferred orientation is present in the crushed crystals.}
\label{sfig1}
\end{figure}

\begin{table}[th]
\centering
\begin{tabular}{cccccc}
\hline
\multicolumn{6}{c}{\multirow{2}{*}{FeI$_2$, $P\bar{3}m1$}}                                                                                                                                                                         \\
\multicolumn{6}{c}{}                                                                                                                                                                         
 \\ \hline\hline
\multicolumn{1}{c||}{\multirow{2}{*}{Radiation}}                                   & \multicolumn{5}{c}{$50\%$, Cu $K\alpha_1$, $\lambda=1.540598$\,\AA}                                     \\
\multicolumn{1}{c||}{}                                                             & \multicolumn{5}{c}{$50\%$, Cu $K\alpha_2$, $\lambda=1.544426$\,\AA}                                     \\ \hline
\multicolumn{1}{c||}{$T$}                                                        & \multicolumn{5}{c}{300~K}                                                                                                       \\ \hline\hline
\multicolumn{1}{c||}{\begin{tabular}[c]{@{}c@{}}Lattice\\ parameters\end{tabular}} & \multicolumn{5}{c}{$a=b=4.05012$~\AA, $c=6.75214$~\AA}                                                                              \\ \hline
\multicolumn{1}{c||}{$B_{\text{iso}}$}                                             & \multicolumn{5}{c}{$1.4032$~\AA$^2$}                                                                                                    \\ \hline\hline
\multicolumn{1}{c||}{Atom}                                                         & \multicolumn{1}{c|}{x}       & \multicolumn{1}{c|}{y}       & \multicolumn{1}{c|}{z}       & \multicolumn{1}{c|}{Occ.}  & Site \\ \hline
\multicolumn{1}{c||}{Fe}                                                           & \multicolumn{1}{c|}{0.00000} & \multicolumn{1}{c|}{0.00000} & \multicolumn{1}{c|}{0.00000} & \multicolumn{1}{c|}{1.000} & 1a   \\ \hline
\multicolumn{1}{c||}{I}                                                            & \multicolumn{1}{c|}{0.33333} & \multicolumn{1}{c|}{0.66667} & \multicolumn{1}{c|}{0.25000} & \multicolumn{1}{c|}{1.000} & 2d   \\ \hline
\end{tabular}
\caption{ Structural parameters determined from Rietveld refinement of powder X-ray diffraction data. }
\end{table}
\clearpage

\section{Data symmetrization}\label{Sec.S2}

We can gain several factors more statistics by exploiting the $\bar{3}m$ Laue symmetry of system to symmetrize the raw neutron scattering data. Elements of symmetry operation are tabulated on the Bilbao crystallographic server under ``General Positions of three-dimensional crystallographic point groups" \cite{si-aroyo2011crystallography}. There are three different coordinate systems for $\bar{3}m$ point group: ``-31m hexagonal axes", ``-3m1 hexagonal axes" and ``-3m rhombohedral axes". Group elements in ``-3m1 hexagonal axes" agree with those in the International Tables for Crystallography, which has its origin at the center of the unit-cell. The ``-31m hexagonal axes" with origin at the corner of the unit-cell is what we need for the symmetrization process working in the environment of Horace \cite{si-ewings2016horace}. The data is stored in the global Cartesian frame in Horace, so we need to transform it to the fractional coordinate before applying the symmetry operations. These group elements are written for a unit-cell with $\gamma = 120^{\circ}$, so this will also be our convention for the unit-cell in the reciprocal space in presenting the data, see Fig.~S8. In the ordered phase, there are three magnetic domains related by 120$^{\circ}$ rotation with somewhat different populations. The symmetrization process averages over all three domains.
\begin{figure}[h!]
\centering
  \includegraphics[width=1\textwidth]{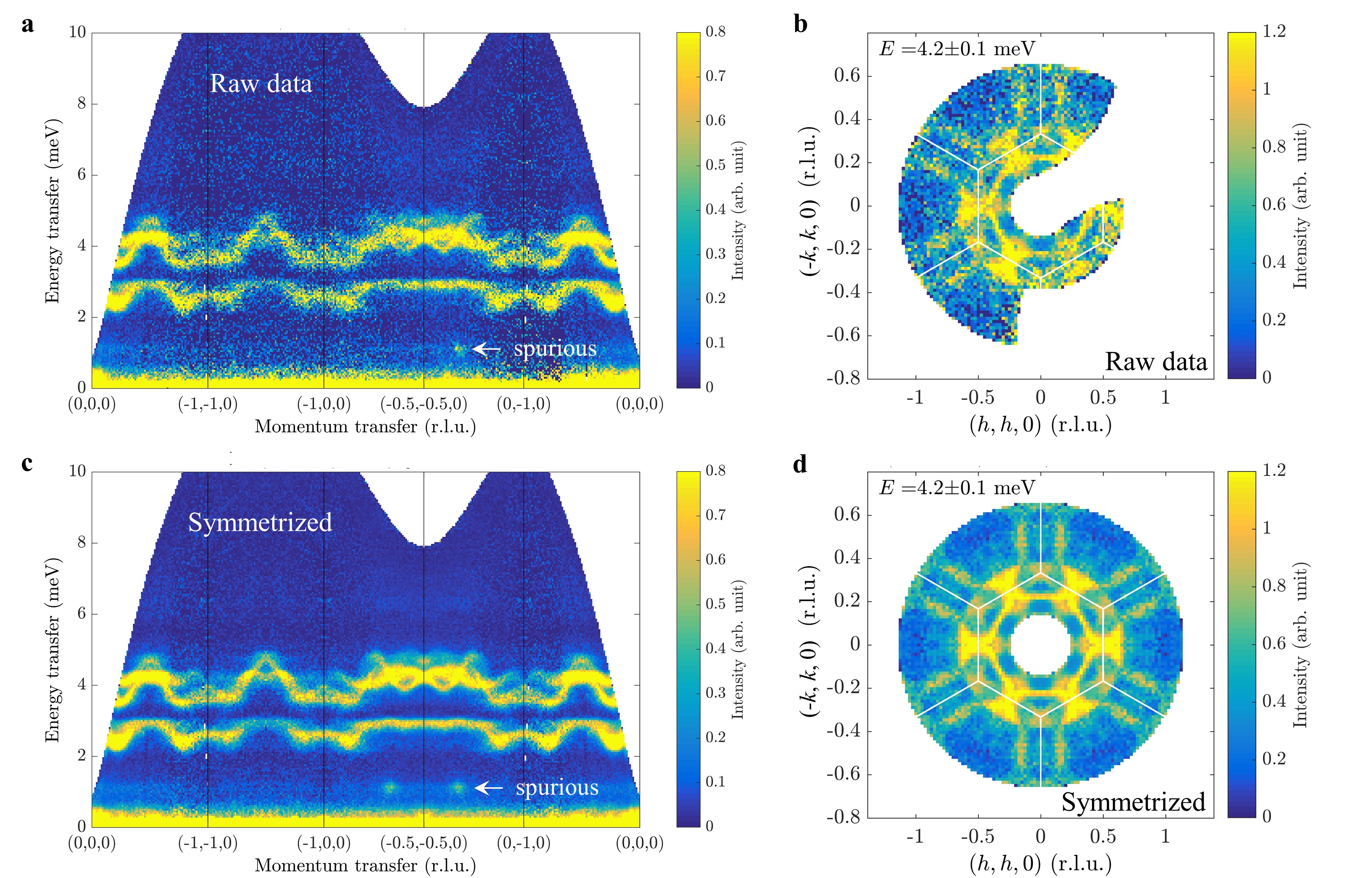}
  \caption{Comparison between the raw data ({\bf a}, {\bf b}) and the symmetrized data ({\bf c}, {\bf d}). The white lines are Brillouin zone boundaries of triangular lattice. }
\label{sfig2}
\end{figure}

\section{Elastic cuts of neutron-scattering data}\label{Sec.S3}

\begin{figure}[th!]
\centering
  \includegraphics[width=1\textwidth]{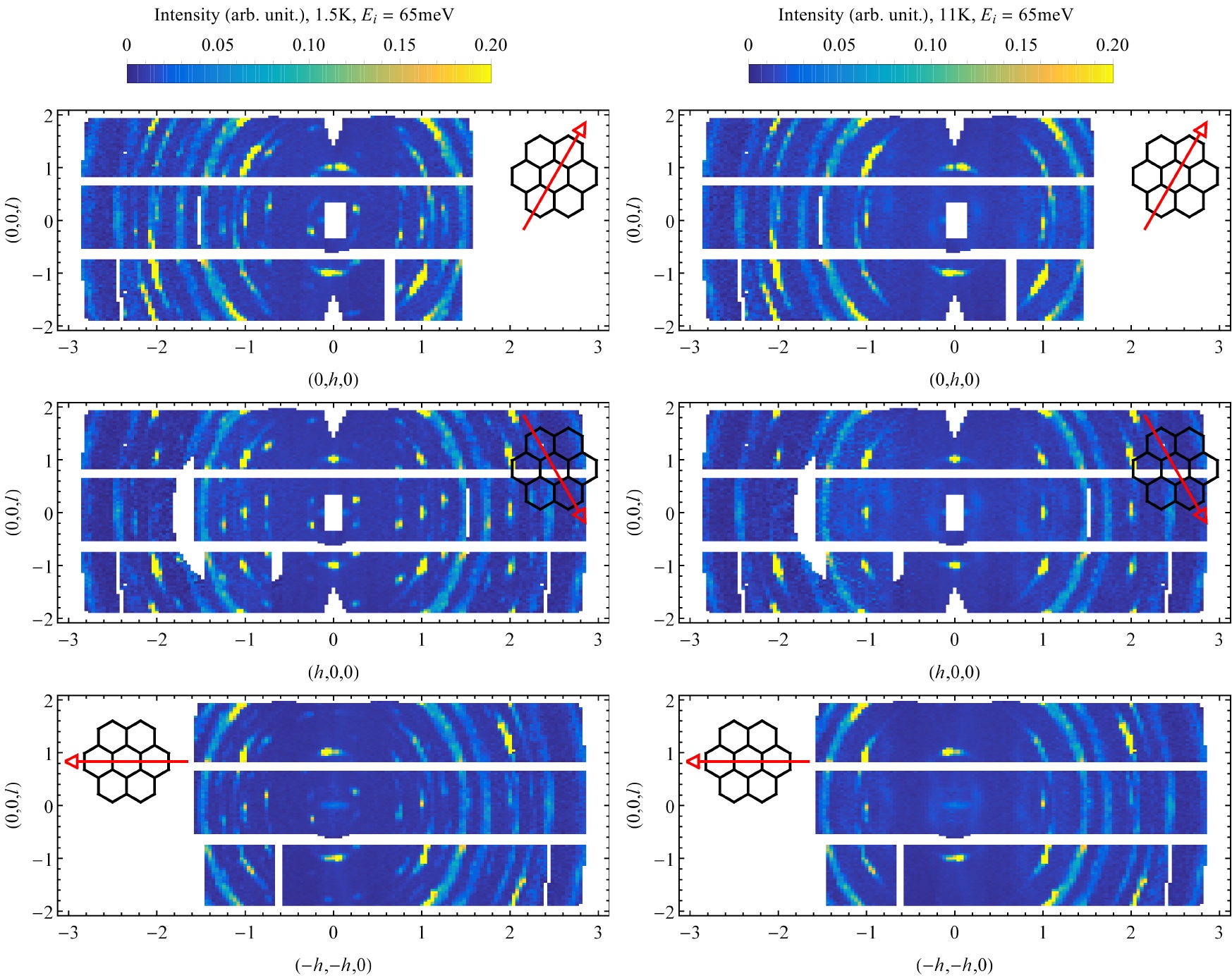}
  \caption{Elastic cuts of neutron-scattering raw data collected at 1.5K (left) and 11K (right) with incoming energy of 65meV. Three symmetry-equivalent cuts related by $120^{\circ}$ rotation are shown for each dataset. The ring-like signals are from aluminum sample holder. Intensities at $(1,0,1)$ and equivalent positions are coming from structural domains which are $60^{\circ}$ rotation from majority of the crystal. The fraction of minority domains estimated from integrated intensities is less than a few percent. Positions of magnetic Bragg peaks are consistent with reported propagation vectors $\textbf{k} = (1/4,0,1/4), (0,1/4,1/4)$ and $(-1/4,-1/4,1/4)$ \cite{si-wiedenmann1988neutron}, corresponding to three magnetic domains of the stripe ordering related by 120 degree rotation. The weak signals at $(1/4,0,-1/4)$ and related positions are the magnetic Bragg peaks from the minority structural domains.}
\label{sfig3}
\end{figure}
\clearpage

\section{Ground-state constraints of exchange parameters}\label{Sec.S4}
\begin{figure}[th!]
\centering
  \includegraphics[width=0.8\textwidth]{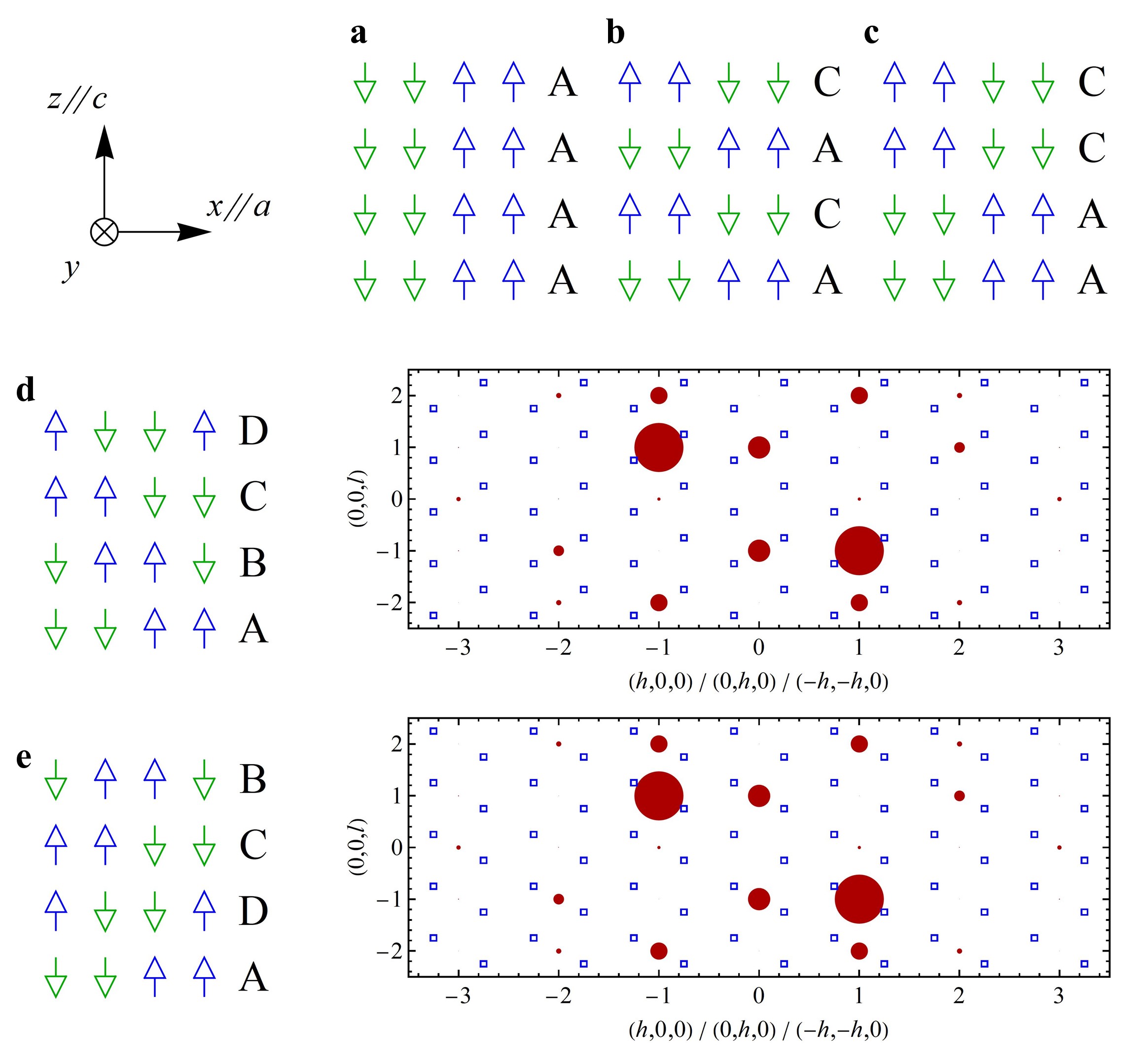}
  \caption{Various low-energy 3D stacking patterns with periodicity less than or equal to four. Positions of magnetic (blue squares) and nuclear Bragg peaks (red disk) of the ABCD and ADCB stacking are shown to the right of respective spin structures. The size of nuclear Bragg peaks is drawn in the scale of their relative intensities. By comparing with experimentally observed diffraction pattern, we find the ABCD stacking is realized in the FeI$_2$. The energies of ABCD, AACC and ADCB stacking do not depend on $J^{'}_{0}$ and $J^{'}_{1}$ and they would be degenerate if $J^{'}_{2a}$ and $J^{'}_{2b}$ were treated the same. 
}
\label{sfig4}
\end{figure}
\noindent The energies of these configurations are 
\begin{align*}
&\text{{\bf a.} AAAA,} \quad E = J_1-J_2-J_3+J_{0}^{'}+2J_{1}^{'}-J_{2\text{a}}^{'}-J_{2\text{b}}^{'}\,,\\
&\text{{\bf b.} ACAC,} \quad E = J_1-J_2-J_3-J_{0}^{'}-2J_{1}^{'}+J_{2\text{a}}^{'}+J_{2\text{b}}^{'}\,,\\
&\text{{\bf c.} AACC,} \quad E = J_1-J_2-J_3\,,\\
&\text{{\bf d.} ABCD,} \quad E = J_1-J_2-J_3-2J_{2\text{a}}^{'}+2J_{2\text{b}}^{'}\,,\\
&\text{{\bf e.} ADCB,} \quad E = J_1-J_2-J_3+2J_{2\text{a}}^{'}-2J_{2\text{b}}^{'}\,.
\end{align*} 
\clearpage

\section{Self-consistent Gaussian approximation}\label{Sec.S5}
\begin{figure}[th!]
\centering
  \includegraphics[width=1\textwidth]{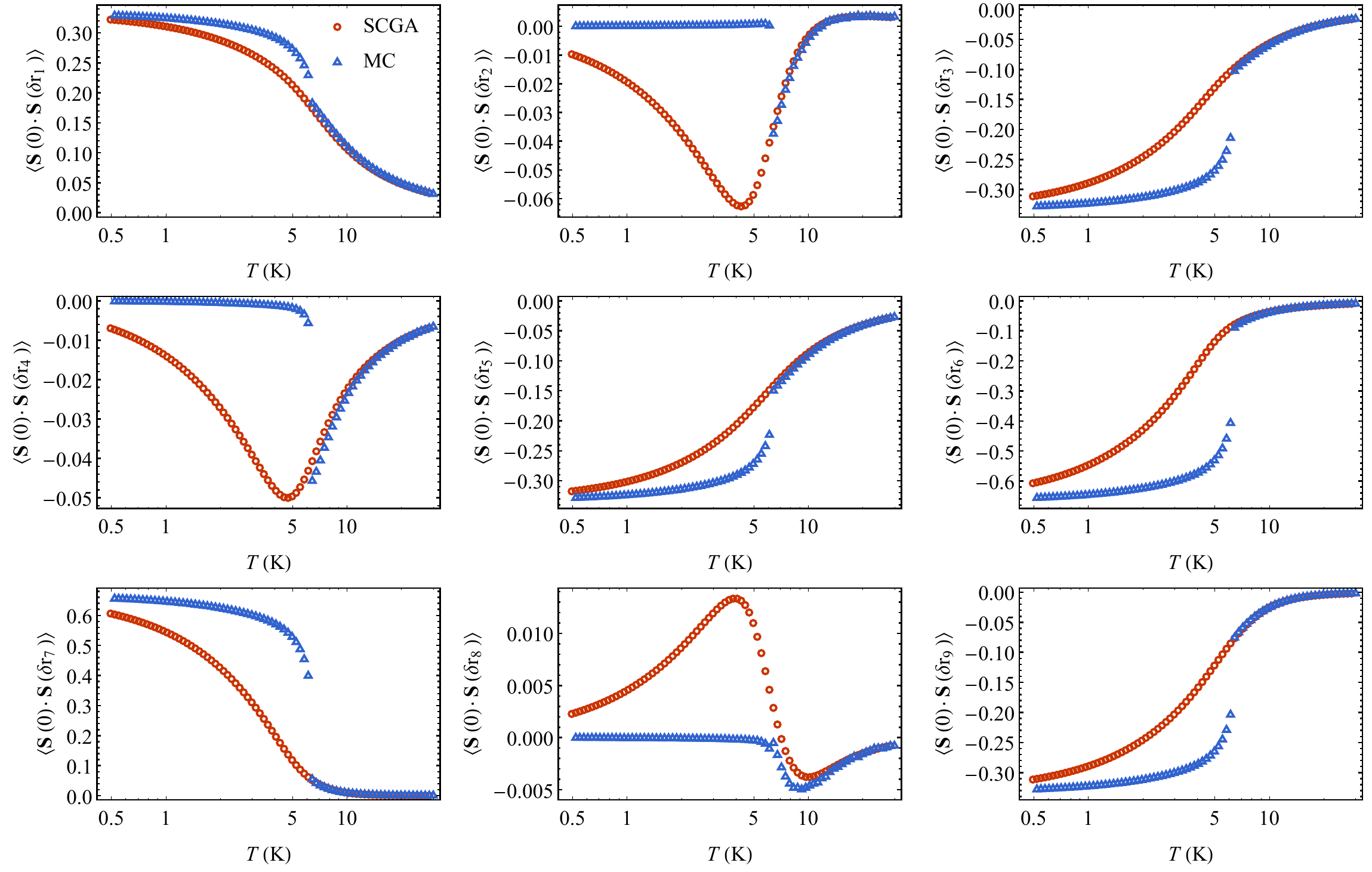}
  \caption{Comparison between SCGA calculation (blue triangles) with MC simulation (red circles) of SU(2) spins for the best fitting parameters of diffuse-scattering data. The good agreement is found for all spin correlations up to the phase transition.}
\label{sfig5}
\end{figure}

\begin{table}[h]
\centering
\begin{tabular}{c|c|c|c|c|c|c|c|c}
\hline
$J_1$~(K) & $J_2$ & $J_3$ & $J^{'}_{0}$ & $J^{'}_{1}$ & $J^{'}_{2a}$ & $D$  & scale & const. bk \\ \hline
-2.74 & 1.31  & 2.46  & -0.41       & 0.59        & 0.82         & 25.2  & 1.96 & 0.19\\ \hline
\end{tabular}
\caption{The best SCGA fitting parameters of a Heisenberg model with single-ion anisotropy for the diffuse-scattering data collected at $T=$11~K.}
\end{table}

Self-consistent Gaussian approximation (SCGA) provides a quantitatively accurate approximation to Monte Carlo simulation of SU(2) spins in the paramagnetic regime \cite{si-conlon2010absent}. It is a very useful tool for extracting exchange parameters from fitting single crystal diffuse scattering data, applicable for both isotropic \cite{si-bai2019magnetic} and anisotropic systems \cite{si-plumb2019continuum}. The basic idea is to relax the hard constraint on fixed spin-length and allow individual spins to fluctuate,
%
\begin{align}
\mathcal{Z} &= \int \prod_{i} d\bS_{i}\delta(\bS_{i}^2-1)\exp\left(-\beta \ham\right)\\
& \approx \int_{-\infty}^{+\infty} \prod_{i,\mu} dS_{i}^{\mu}\exp\left(-\dfrac{1}{2}\lambda (S_{i}^{\mu})^2\right)\exp\left(-\beta \ham\right),
\end{align}
while introducing a Langrange multiplier $\lambda$ is to ensure the length of the spin is 1 on average
\begin{align}
\dfrac{1}{N}\sum_{i\mu}\left\langle (S_{i}^{\mu})^{2}\right\rangle = 1,
\end{align}
{where $N$ is the number of lattice sites.}
%
%
%
%

It is convenient to work in the momentum space by {a Fourier transform} of the Hamiltonian 
{\begin{align}
\ham & = \dfrac{1}{2}\sum_{n}\sum_{i,j}\sum_{\mu,\nu}\sum_{\alpha,\beta} A_{i\alpha,j\beta}^{n}J^{\mu\nu}_{n} S^{\mu}_{\alpha,\br_i} S^{\nu}_{\beta,\br_j} \\
& =\dfrac{1}{2}\sum_{n}\sum_{i,j}\sum_{\mu,\nu}\sum_{\alpha,\beta}\dfrac{1}{N_{\text{uc}}}\sum_{\bq,\bp}\, e^{i(\bR_i+\bd_{\alpha})\cdot\bq}e^{i(\bR_j+\bd_{\beta})\cdot\bp}A_{i\alpha,j\beta}^{n}J^{\mu\nu}_{n}  S^{\mu}_{\alpha,\bq} S^{\nu}_{\beta,\bp}\\
& =\dfrac{1}{2}\sum_{n}\sum_{i}\sum_{\mu,\nu}\sum_{\alpha,\beta}\dfrac{1}{N_{\text{uc}}}\sum_{\bq,\bp}\, e^{i(\bR_i+\bd_{\alpha})\cdot\bq}e^{i(\bR_i+\bd_{\alpha})\cdot\bp}J^{\mu\alpha,\nu\beta}_{n}(\bp)  S^{\mu}_{\alpha,\bq} S^{\nu}_{\beta,\bp}\\
& =\dfrac{1}{2}\sum_{n}\sum_{\mu,\nu}\sum_{\alpha,\beta}\sum_{\bp}\,J_{n}^{\mu\alpha,\nu\beta}(\bp)  S^{\mu}_{\alpha,-\bp} S^{\nu}_{\beta,\bp}\\
& = \dfrac{1}{2}\sum_{n}\sum_{\bp}\bS_{-\bp}^{T}\cdot\bJ_n(\bp)\cdot\bS_{\bp},
\end{align}}
where $\mu,\nu = x,y,z$ labels spin components, $\alpha,\beta$ sublattice index and $i,j$ unit-cell index, {${\bR_i}$, $\br_i$ ($\equiv \bR_i + \bd_{\alpha}$) the coordinate of $i$-th unit-cell and lattice site}. $A_{i\alpha,j\beta}^{n}$ is the adjacency matrix of the {$n$-th} neighbor. The Fourier transform of the interaction matrix, spins and the $\delta$-funciton are used 
{\begin{align}
&J_{n}^{\mu\alpha,\nu\beta}(\bp) = J^{\mu\nu}_{n}A_{\alpha\beta}^{n}(\bp)= J^{\mu\nu}_{n}\sum_{j}A_{i\alpha,j\beta}^{n}\exp(-i\bp\cdot(\bR_i-\bR_j+\bd_{\alpha}-\bd_{\beta}))),\\
&S^{\mu}_{\alpha,\bq} = \dfrac{1}{\sqrt{N_{\text{uc}}}}\sum_{i}\, S^{\mu}_{\alpha,\br_i} e^{-i \bq\cdot (\bR_i+ \bd_{\alpha})},\\
&S^{\mu}_{\alpha,\br_i} = \dfrac{1}{\sqrt{N_{\text{uc}}}}\sum_{\bq} S^{\mu}_{\alpha,\bq} e^{i \bq\cdot (\bR_i+ \bd_{\alpha})},\\
&\delta(\bq-\bp) = \dfrac{1}{N_{\text{uc}}}\sum_{i} e^{-i\bR_i\cdot(\bq-\bp)},
\end{align}}
{where $N_{\text{uc}}$ is the number of crystallographic unit-cell.}
The partition function can be written in a Gaussian form
\begin{align}
\mathcal{Z} = \int \prod_{\bq}d\bS(-\bq)d\bS(\bq) \exp(-\dfrac{1}{2}\bS(-\bq)\cdot\left(\lambda {\bf{1}} +\beta \sum_{n}\bJ_n(\bq)\right)\cdot \bS(\bq)^{T}),
\end{align}
where $\bS(\bq)$ is a $3N_{\text{basis}}$-component vector and $\bJ_n(\bq)$ is a $3N_{\text{basis}}\times 3N_{\text{basis}}$ matrix, {$N_{\text{basis}}$ the number of atom in the crystallographic unit-cell}. The spin correlation is given by 
{\begin{align}
S^{\mu\nu}(\bq) = \sum_{\alpha,\beta}\left\langle S^{\mu}_{\alpha,-\bq}S^{\nu}_{\beta,\bq}\right\rangle = \sum_{\alpha,\beta}\left(\left[\lambda {\bf{1}} +\beta \sum_{n}\bJ_n(\bq)\right]^{-1}\right)^{\mu\nu}_{\alpha\beta}\,.
\end{align}}
The Lagrangian multiplier $\lambda$ is solved from 
\begin{align}
1 = \dfrac{1}{N_{\text{uc}}}\sum_{\bq,\mu}S^{\mu\mu}(\bq)
= \dfrac{1}{N_{\text{basis}}N_{\text{uc}}}\sum_{\bq}\text{tr}\left[\lambda {\bf{1}} +\beta \sum_{n}\bJ_{n}(\bq)\right]^{-1}\,.
\end{align}
The equal-time structure factor observed in unpolarized neutron scattering is 
\begin{align}
S(\bQ) = \sum_{\mu\nu}\left(\delta^{\mu\nu} -\dfrac{ Q^{\mu}Q^{\nu}}{Q^2}\right)S^{\mu\nu}(\bQ),
\end{align}
where $\bQ = {\bf G} +\bq$ and $\bf G$ is a reciprocal lattice vector. 
Spin correlations in real space can be calculated as well
{\begin{align}
\langle S^{\mu}_{\br} S^{\nu}_{\br+\delta\br_{n}} \rangle &=\dfrac{1}{M_{n}}\dfrac{1}{N_{\text{basis}}}\dfrac{1}{N_{\text{uc}}}\sum_{i,j}\sum_{\alpha,\beta}A_{i\alpha,j\beta}^{n}\langle S^{\mu}_{\bR_i+\bd_{\alpha}} S^{\nu}_{\bR_j+\bd_{\beta}}\rangle\\
&=\dfrac{1}{M_{n}}\dfrac{1}{N_{\text{basis}}}\dfrac{1}{N_{\text{uc}}}\sum_{\mu,\nu}\sum_{\bp}\, A^{n}_{\alpha\beta}(\bp)\langle S_{\alpha,-\bp}^{\mu} S^{\nu}_{\beta,\bp}\rangle\,.
\end{align}}
where $M_n$ is the number of {$n$-th} neighbor.
The single-ion term, {$-D\sum_{i\alpha}({S}_{\alpha,\br_i}^{z})^{2}$}, after Fourier transform,  adds a constant $-2D$ to the $zz$ component for each sublattice in the interaction matrix.

\clearpage
\section{Generalized spin-wave theory}\label{Sec.S6}

The FeI$_{2}$ compound is described by the effective $S=1$ spin model, 
\begin{eqnarray}
{\cal H}  =  \sum_{\langle ij\rangle}\sum_{\mu\nu}{S}_{i}^{\mu}{\cal J}_{ij}^{\mu\nu}{S}_{j}^{\nu}
- D\sum_{i} Q^{zz}_{ i},\label{eq:hamiltnian}
\end{eqnarray}
where ${S}_i^{\mu}, \mu=x,y,z$ is the spin-$1$ operator and the single-ion anisotropy term is proportional to the $(zz)$ component of quadrupolar moment $Q^{\mu \nu}_i=({S}_i^{\mu} {S}_i^{\nu} + {S}_i^{\nu} {S}_i^{\mu})/2 -2/3 {{\delta^{\mu\nu}} }$ (symmetric traceless components of ${{\bm S}}_i \otimes {{\bm S}}_i$). 
The spin-exchange tensor ${\cal J}_{ij}^{\mu\nu}$ 
is described in the main text and a detailed symmetry analysis is provided in S.8.

The distinctive property of FeI$_{2}$ is that its low-energy modes include both dipolar and quadrupolar fluctuations because of an unusual balance between the magnitude of the exchange interaction and the single-ion anisotropy ($D$ is comparable to $zJ$, where $z=6$ is the coordination number). This observation indicates that we need to generalize the SU(2) spin-wave theory to include both types of low-energy modes. The three components of the magnetization and the five components of the quadupolar moment generate the {SU(3) unitary} transformations in the 3-dimensional Hilbert space of {a} $S=1$ spin.
Correspondingly, an SU(3) spin-wave theory can simultaneously account for the low-energy dipolar and quadrupolar fluctuations of FeI$_{2}$. This generalization can be implemented by  introducing the   SU(3) Schwinger boson representation of the spin operators
${S}_{i}^{\mu}={\psi}_i^{\dagger} L^{\mu}{\psi}_i$, where ${\psi}_i=(b_{i,+1},b_{i,0},b_{i,-1})$
and
\begin{eqnarray}
L^x = 
\left(\begin{array}{ccc}
0 & -{i\over \sqrt{2}} & 0\\
{1\over \sqrt{2}} & 0 & {1\over \sqrt{2}}\\
0 & {1\over \sqrt{2}} & 0
\end{array}\right),
L^y = 
\left(\begin{array}{ccc}
0 & {1\over \sqrt{2}} & 0\\
{i\over \sqrt{2}} & 0 & -{i\over \sqrt{2}}\\
0 & {i\over \sqrt{2}} & 0
\end{array}\right),
L^z = 
\left(\begin{array}{ccc}
1 & 0 & 0\\
0 & 0 & 0 \\
0 & 0 & -1
\end{array}\right).
\end{eqnarray}
The number of Schwinger bosons per site,  $\sum_{m}  b_{i m}^{\dagger}b_{i m}  =  2S=1$ is determined by the spin size $S=1$.

The observed magnetic order in FeI$_{2}$ is described by a condensation
of bosons in the single-particle state 
\begin{eqnarray}
\rvert\psi_{i}\rangle & = & \left(\begin{array}{c}
e^{i\alpha_{1}(i)}\cos[\phi(i)]\sin[\theta(i)]\\
e^{i\alpha_{2}(i)}\sin[\phi(i)]\sin[\theta(i)]\\
\cos[\theta(i)]
\end{array}\right)
\end{eqnarray}
on the basis of $\{\rvert i,+1\rangle,\rvert i,0 \rangle,\rvert i,-1 \rangle\}$,
where $\rvert i,m\rangle=b_{i,m}^{\dagger}\rvert \emptyset  \rangle$ with $\rvert \emptyset \rangle$
being the vacuum of  Schwinger {bosons.} The parameters $\theta(i),\phi(i),\alpha_{1}(i),\alpha_{2}(i)$
are determined by minimizing the mean field energy $(\prod_{i=1}^N \langle \psi_{i} \rvert) {\cal H} (\prod_{i=1}^N \rvert\psi_{i}\rangle)$ with $N$ the number of 
lattice sites. The optimal state $\rvert\psi_{i}\rangle$ is a SU(3) coherent state, which in general cannot be
obtained from the \textit{fully polarized} state $(1,0,0)^{T}$ by applying a SU(2) rotation.
{Fig.~\ref{sfig6}{\bf a}} shows the length of dipole
moment $d=\sqrt{\sum_{\mu}\langle {S}_{i}^{\mu}\rangle^{2}}$ {
in} the parameter space $(\theta,\phi)$ for optimal values of $\alpha_{1,2}$.
We note that $d=1$ for an SU(2) spin coherent state, which is indicated with a dark blue line in Fig.~\ref{sfig6}.
{It turns out that the optimal state of FeI$_2$ deviates from the SU(2) spin coherent state. { This deviation}
arises from the $J_{1}^{z\pm}$ term that breaks the axial symmetry about the $c$-axis. The net result is a 
$10^{o}$ canting of the magnetic moments away from the $c$-axis and also a small reduction in the magnitude of the moment. 
Nevertheless, { the magnitude of the moment is very close 1,} as anticipated from the rather strong easy-axis single-ion ansiotropy. Therefore, the optimal SU(3) state is practically close to the SU(2) state.}

\begin{figure}[t]
\centering
\includegraphics[width=0.9\textwidth]{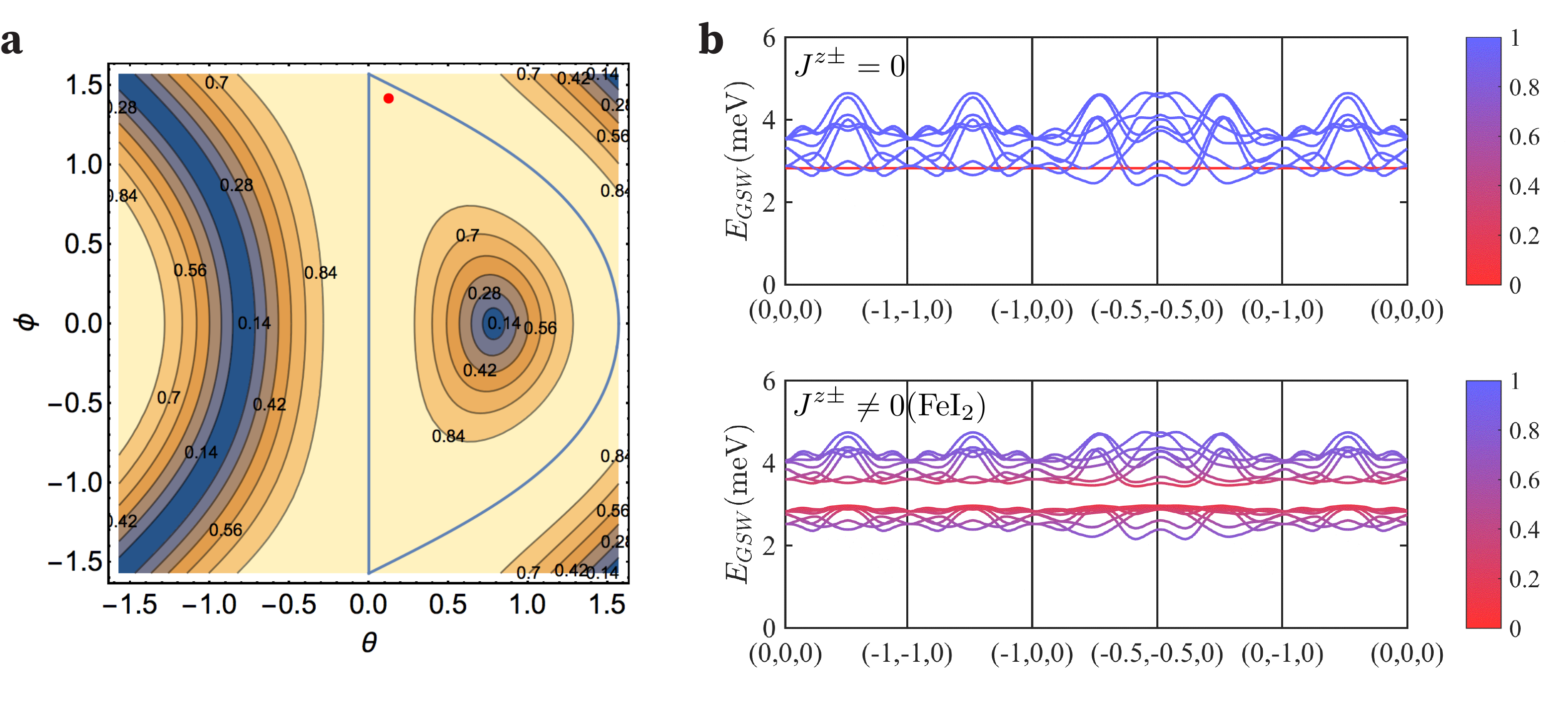}
\caption{\textbf{a.} Length of dipole moment $d$ in the parameter space $(\theta,\phi)$. The dark blue line indicates 
the manifold of SU(2) spin coherent states characterized by $d=1$. The red dot indicates the  optimal mean field state for 
FeI$_2$. \textbf{b.} Excitation spectrum obtained from the SU(3) spin wave theory before and after including $J_{z\pm}$. Color scale: blue
and red colors refer to the weights of the $\beta_{i,0}$ and $\beta_{i,-1}$ bosons, respectively. }
\label{sfig6}
\end{figure}

To compute the excitation spectrum, it is convenient to work in the local reference frame defined by
the SU(3) rotation
\begin{equation}
\left(\begin{array}{c}
\beta_{i,+1}\\
\beta_{i,0}\\
\beta_{i,-1}
\end{array}\right)=U_{i}^{\dagger}\cdot\left(\begin{array}{c}
b_{i\uparrow}\\
b_{i0}\\
b_{i\downarrow}
\end{array}\right),\label{eq:su3_rotation-1}
\end{equation}
where $U_{i}\equiv A_{1}\cdot A_{2}$ with
\begin{eqnarray}
A_{1} & \equiv & \left(\begin{array}{ccc}
\sin(\theta)\cos(\phi)e^{i\alpha_{1}} & \cos(\theta)\cos(\phi)e^{i\alpha_{1}} & -\sin(\phi)e^{-i(\alpha_{2}+\alpha_{3})}\\
\sin(\theta)\sin(\phi)e^{i\alpha_{2}} & \cos(\theta)\sin(\phi)e^{i\alpha_{2}} & \cos(\phi)e^{-i(\alpha_{1}+\alpha_{3})}\\
\cos(\theta)e^{i\alpha_{3}} & -\sin(\theta)e^{i\alpha_{3}} & 0
\end{array}\right),
\end{eqnarray}
and
\begin{eqnarray}
A_{2} & \equiv & \left(\begin{array}{ccc}
1 & 0 & 0\\
0 & \cos(\chi)e^{-i\beta_{1}} & \sin(\chi)e^{i(\beta_{2}-\alpha_{1}-\alpha_{2}-\alpha_{3})}\\
0 & -\sin(\chi)e^{-i(\beta_{2}-\alpha_{1}-\alpha_{2}-\alpha_{3})} & \cos(\chi)e^{i\beta_{1}}
\end{array}\right).
\end{eqnarray}
Up to a phase factor, the first column of $U$ is the optimal state 
$\rvert\psi_{i}\rangle=\beta_{i,+1}^{\dagger}\rvert \emptyset \rangle$.
The other two flavors (m=0,-1) of $\beta_{i,m}$ are defined by
the second and third columns of $U_{i}$, which are orthogonal to the first column. 
For $U_{i}\in SU(2)$,
the boson $\beta_{i,m}$ carries a quantized angular momentum $m=1$
along the quantization axis $\hat{{\bm n}}_{i}=\langle\psi_{c}\rvert {{\bm S}}_{i}\rvert\psi_{c}\rangle$.
The condensation of the $\beta_{i,+1}$ boson gives rise to the local dipople moments along the direction $\hat{{\bm n}}_{i}$. The $\beta_{i,-1}$ boson changes the angular momentum by $-2$ relative to the local 
quantization axis, {\it i.e.}, it { generates} a local  quadrupolar fluctuation. 
Unless there is a finite hybridyzation between the $\beta_{i,0}$ and $\beta_{i,-1}$ bosons, quadrupolar { excitations} 
remain invisible to inelastic neutron scattering.
We stress that, although the rotation matrix $U_{i}$ can be approximated by an SU(2) matrix in FeI$_{2}$,
the normal modes include a rather strong hybridyzation between the $\beta_{i,0}$ and $\beta_{i,-1}$ bosons, which enables the observation of predominantly quadrupolar excitations with inelastic neutron scattering.

Since the $\beta_{i,+1}$ boson is macroscopically occupied, $\langle\beta_{i,+1}\rangle=\langle\beta_{i,+1}^{\dagger}\rangle\simeq\sqrt{M}$ ($M=1$ for the case under consideration), we  assume that $\langle\beta_{i,0}^{\dagger}\beta_{i,0}\rangle,\langle\beta_{i,-1}^{\dagger}\beta_{i,-1}\rangle \ll M$. This assumption justifies an expansion in  the {
small parameter $1/M$}:
\begin{eqnarray}
\beta_{i,+1} & = & \beta_{i,+1}^{\dagger}=\sqrt{M-\beta_{i,0}^{\dagger}\beta_{i,0}-\beta_{i,-1}^{\dagger}\beta_{i,-1}}\nonumber \\
 & \simeq & \sqrt{M}\left(1-\frac{1}{2M}\beta_{i,0}^{\dagger}\beta_{i,0}-\frac{1}{2M}\beta_{i,-1}^{\dagger}\beta_{i,-1}+{\cal Q}(\frac{1}{(M)^{2}})\right).
\end{eqnarray}
Consequently, we have the semi-classical expansion of the dipolar and
quadrupolar operators
\begin{eqnarray}
{S}_{i}^{\mu} & = & M{\cal S}_{c}^{\mu}(i)+\sqrt{M}\sum_{m\neq1}\left({\cal S}_{1m}^{\mu}(i)\beta_{i,m}+h.c.\right)+\sum_{m,n\neq1}{\cal S}_{mn}^{\mu}(i)\beta_{i,m}^{\dagger}\beta_{i,n}+{\cal O}(\frac{1}{\sqrt{M}}),\label{eq:spin}
\end{eqnarray}
\begin{eqnarray}
{Q}^{zz}_i & = & M{\cal Q}_{c}^{zz}(i)+\sqrt{M}\sum_{m\neq1}\left({\cal Q}_{1m}^{zz}(i)\beta_{i,m}+h.c.\right)+\sum_{m,n\neq1}{\cal Q}_{mn}^{zz}(i)\beta_{i,m}^{\dagger}\beta_{i,n}+{\cal O}(\frac{1}{\sqrt{M}}),\label{eq:quadrupole}
\end{eqnarray}
where
\begin{eqnarray}
{\cal S}_{c}^{\mu}(i) &=& \left(U_i^{\dagger}L^{\mu}U_i\right)_{11},  {\cal Q}_{c}^{zz}(i)=\left(U_i^{\dagger}O^{zz}U_i\right)_{11},\label{eq:ob1} \\
{\cal S}_{1m}^{\mu}(i)&=&\left(U_i^{\dagger}L^{\mu}U_i\right)_{1m}, {\cal Q}_{1m}^{zz}(i)=\left(U_i^{\dagger}O^{zz}U_i\right)_{1m},\label{eq:ob2} \\
{\cal S}_{mn}^{\mu}(i) & = & \left(U_i^{\dagger}L^{\mu}U_i\right)_{mn}-\left(U_i^{\dagger}L^{\mu}U_i\right)_{11}\delta_{mn},\label{eq:ob3}\\
{\cal Q}_{mn}^{zz}(i) & = & \left(U_i^{\dagger}O^{zz}U_i\right)_{mn}-\left(U_i^{\dagger}O^{zz}U_i\right)_{11}\delta_{mn},\label{eq:ob4}
\end{eqnarray}
with {$O^{zz}=(L^z)^2$}. Note that the variables defined in Eqs. (\ref{eq:ob1}-\ref{eq:ob4})
depend only on the sublattice index because of the translation symmetry of the magnetic structure. 
Applying the above formula, we obtain a generalized semi-classical expansion of the spin Hamiltonian
\begin{eqnarray}
{\cal H} & = & {\cal E}^{(0)}+{\cal H}^{(2)}+{\cal O}(M^{0}),
\end{eqnarray}
where
\begin{eqnarray}
{\cal E}^{(0)} & = & M^{2}\sum_{\langle ij\rangle}\sum_{\mu\nu}{\cal S}_{c}^{\mu}(i){\cal J}_{ij}^{\mu\nu}{\cal S}_{c}^{\nu}(j)-D\sum_{i} ({\cal Q}_{c}^{zz}(i)+{2\over3}),
\end{eqnarray}
and
\begin{eqnarray}
{\cal H}^{(2)} & = & \frac{1}{2}\sum_{{\bm q}\alpha\beta}\sum_{m,n\neq1}\left(\begin{array}{c}
\beta_{(\alpha,\bm{q})m}^{\dagger}\\
\beta_{(\beta,\bm{q})m}^{\dagger}\\
\beta_{(\alpha,\bar{\bm{q}})m}\\
\beta_{(\beta,\bar{\bm{q}})m}
\end{array}\right)^{T}\left(\begin{array}{cccc}
\Delta_{mn}^{\alpha} & \Theta_{mn;\bm{q}}^{\alpha\beta} & 0 & \Xi_{mn;\bm{q}}^{\alpha\beta}\\
\Theta_{nm;\bm{q}}^{\alpha\beta*} & \Delta_{mn}^{\beta} & \Xi_{nm;\bar{\bm{q}}}^{\alpha\beta} & 0\\
0 & \Xi_{mn;\bar{\bm{q}}}^{\alpha\beta*} & \Delta_{nm}^{\alpha} & \Theta_{mn;\bar{\bm{q}}}^{\alpha\beta*}\\
\Xi_{nm;\bm{q}}^{\alpha\beta*} & 0 & \Theta_{nm;\bar{\bm{q}}}^{\alpha\beta} & \Delta_{nm}^{\beta}
\end{array}\right)\left(\begin{array}{c}
\beta_{(\alpha,\bm{q})n}\\
\beta_{(\beta,\bm{q})n}\\
\beta_{(\alpha,\bar{\bm{q}})n}^{\dagger}\\
\beta_{(\beta,\bar{\bm{q}})n}^{\dagger}
\end{array}\right)\nonumber \\
 &  & -\sum_{m\neq1}\sum_{i}\Delta_{mm}^{i},
\end{eqnarray}
where $\bar{\bm{q}}\equiv-\bm{q}$, $\alpha=1,...,4$ is the sublattice index, and 
$\beta_{(\alpha,{\bm q})\sigma} = N_{uc}^{-1/2} \sum_{\bm{r}}e^{-i\bm{q}\cdot \bm{r}}\beta_{(\alpha,\bm{r})\sigma}$.
{$N_{uc}$ is the total number of the magnetic unit cells and $\bm{r}$ { denotes the coordinates of each unit cell}}. We now introduce the quantities
\begin{eqnarray}
\Delta_{mn}^{\alpha} & = & M\sum_{\beta}\sum_{   \bm{\delta}_{\alpha\beta}>0  }\sum_{\mu\nu}{\cal S}_{c}^{\mu}(\beta){\cal J}_{{\bm \delta}_{\alpha\beta}}^{\mu\nu}{\cal S}_{mn}^{\nu}(\alpha)  
- {D\over 2}  {\cal Q}_{mn}^{zz}(\alpha),\\
\Theta_{mn;\bm{q}}^{\alpha\beta} & = & M\sum_{\bm{\delta}_{\alpha\beta}>0}\sum_{\mu\nu}{\cal S}_{1m}^{\mu*}(\alpha){\cal J}_{\bm{\delta}_{\alpha\beta}}^{\mu\nu}{\cal S}_{1n}^{\nu}(\beta)e^{i\bm{q}\cdot\bm{\delta}_{\alpha\beta}}, \\
\Xi_{mn;\bm{q}}^{\alpha\beta} & = & M\sum_{\bm{\delta}_{\alpha\beta}>0}\sum_{\mu\nu}{\cal S}_{1m}^{\mu*}(\alpha){\cal J}_{\bm{\delta}_{\alpha\beta}}^{\mu\nu}{\cal S}_{1n}^{\nu*}(\beta)e^{i\bm{q}\cdot\bm{\delta}_{\alpha\beta}},
\end{eqnarray}
where $\bm{\delta}_{\alpha\beta}$ are  the bond vectors connecting
sublattices $\alpha$ and $\beta$ and the summation over $\bm{\delta}_{\alpha\beta}>0$ avoids double-counting of each bond.

To fit the inelastic neutron-scattering data, we compute the dynamical
spin structure factor, $S_{\mu\nu}({\bf q},\omega)=2\Theta(\omega) \chi_{\mu\nu}^{\prime\prime}({\bf q},\omega)$, where $\Theta(\omega)$ is the Heaviside step
function and $\chi_{\mu\nu}^{\prime\prime}({\bf q},\omega)$ is the imaginary part of the dynamical spin susceptibility
{\begin{equation}
i\chi_{\mu\nu}({\bf q},\omega)=\frac{1}{4} \sum_{\alpha\beta}\int_{0}^{\infty} dt e^{i\omega t} \langle [{S}_{\alpha,\bm{q}}^{\mu}(t), {S}_{\beta,-\bm{q}}^{\nu}(0)]\rangle,
\end{equation}}
where {$S_{\alpha,{\bm q}}^{\mu} = N_{uc}^{-1/2} \sum_{\bm{r}}e^{-i\bm{q}\cdot \bm{r}}S_{\alpha,\bm{r}}^{\mu}$}. 
By applying the semiclassical $1/M$ expansion up to the lowest non-trivial order, we
obtain
\begin{eqnarray}
\chi_{\alpha\beta}^{\mu\nu}({\bf q},\omega) =  
M\sum_{mn}
\left(\begin{array}{c}
{\cal S}_{1m}^{\mu}(\alpha)\\
{\cal S}_{m1}^{\mu}(\alpha)
\end{array}\right)^{T}
\left( \begin{array}{cc}
{\cal G}_{(\alpha,m)(\beta,n)}^{(N)}(\bm{q},\omega) & \check{{\cal G}}_{(\alpha,m)(\beta,n)}(\bm{q},\omega) \\
\hat{{\cal G}}_{(\alpha,m)(\beta,n)}(\bm{q},\omega) & {\cal G}_{(\beta,n)(\alpha,m)}^{(N)}(-\bm{q},-\omega)
\end{array} \right)
\left(\begin{array}{c}
{\cal S}_{n1}^{\nu}(\beta)\\
{\cal S}_{1n}^{\nu}(\beta)
\end{array}\right)
\end{eqnarray}
where the single-particle Green's function is given by
\begin{eqnarray}
 \left(\begin{array}{cc}
{\cal G}^{(N)}(\bm{q},\omega) & \check{{\cal G}}(\bm{q},\omega) \\
\hat{{\cal G}}(\bm{q},\omega) & {\cal G}^{(N)}(-\bm{q},-\omega)
\end{array}\right) 
 &=& \left( -(\omega+i0^+)A + {\cal H}^{(2)}\right)^{-1}, \\
A & = & \left(\begin{array}{cc}
I_{8\times8} & 0\\
0 & -I_{8\times8}
\end{array}\right).
\end{eqnarray}
$I_{8\times8}$ is the $8 \times 8$ identity matrix. Fig.~S6\textbf{b} shows the  excitation spectrum obtained from the SU(3) spin wave theory for the Hamiltonian parameters that were obtained by fitting  the inelastic neutron scattering 
data. The relative weight of the $\beta_{i,0}$ and $\beta_{i,-1}$
bosons is shown in blue and red colors, respectively. As expected, the two modes are strongly hybridized in the region where 
the dipolar and quadrupolar excitations have practically the same energy.

\clearpage
\section{GSWT fitting of inelastic neutron-scattering data}

\begin{figure}[h]
\centering
\includegraphics[width = 1\textwidth]{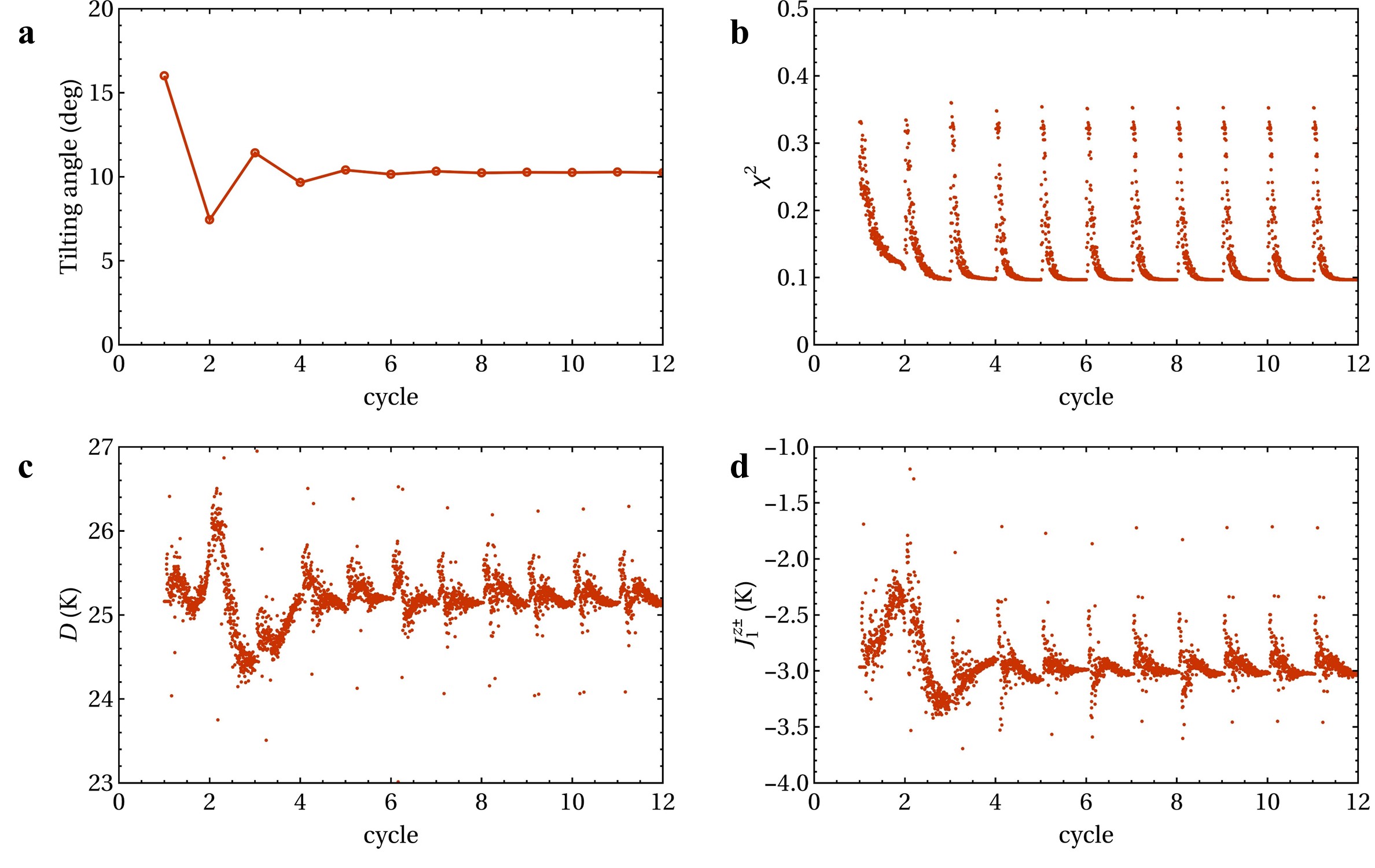}
\caption{Progress of various quantities in the GSWT fittings - the tilting angle ({\bf a}), the reduced $\chi^2$ ({\bf b}), the single-ion anistropy $D$ ({\bf c}) and the off-diagonal exchange $J^{z\pm}_{1}$ ({\bf d}) which is responsible for the hybridization.}
\label{sfig7}
\end{figure}

We perform a pixel-to-pixel fitting to selected cuts of symmetrized inelastic neutron-scattering data along a high-symmetry path in the $(h,k,0)$-plane and four out-of-plane paths, shown in Fig.~S8. The raw data was collected on SEQUOIA at $1.8$~K with incoming neutron energy $E_i=12$~meV. Our anisotropic model includes four exchange constants $\lbrace J^{\pm}_{1},J^{zz}_{1},J^{\pm\pm}_{1},J^{z\pm}_{1}\rbrace$  for NN bonds, two exchange constants $\lbrace J^{\pm},J^{zz}\rbrace$ for each of the other bonds, one single-ion anisotropy, one scaling parameter and one damping parameter, 17 parameters in total. The $J^{zz}_0$ and $J^{zz}_1$ coupling does not have any effect on calculated inelastic spectrum, so they are set to zero in the fitting. The reduced $\chi^2$ defined as $\sum_{i}(I_{\text{obs}}-I_{\text{cal}})/N_{\text{pix}}$ is minimized by varying the other 15 parameters using the Nelder-Mead method implemented in the NLopt package \cite{si-nlopt,si-nelder1965simplex}. The fitting procedure runs in cycles, each containing 300 minimization steps. The ground state is optimized at the beginning of each cycle. The optimized magnetic structure appears to tilt away from c-axis while remains collinear. The tilting angle converges to around $10^{\circ}$. The fitting starts with the best SCGA fitting parameters of diffuse-scattering data with small randomization. We performed 80 independent fittings with randomized starting parameters and computed the standard deviation of converged results after 11 cycles as an estimation of error. The progress for one of the fitting is shown in Fig.~S7.  The best fitting parameters are listed in Tab.~S3. All the $zz$-couplings only contribute constants to the calculated spectrum, which leads to significant parameter dependence among them. This can be seen from their large standard deviations. The spectrum is however extremely sensitive to the combination $E_{\text{SIBS}}=4(-J^{zz}_1+J^{zz}_2+J^{zz}_3+2J_{2a}^{'zz})$ which determines the energy of the single-ion bound state. All the transverse couplings and the single-ion anisotropy $D$ are determined with very small error. 

\begin{figure}[ht]
\centering
\includegraphics[width = 0.5\textwidth]{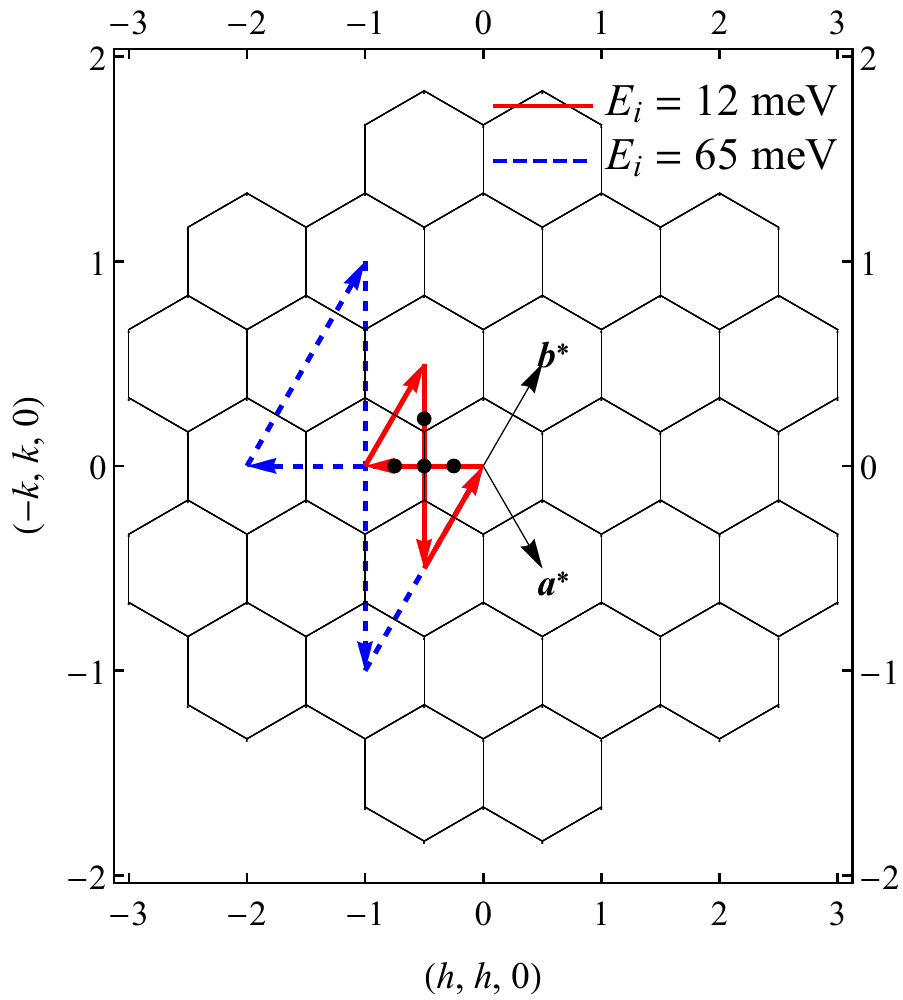}
\caption{The high-symmetry path used in the GSWT fitting and the convention of reciprocal lattice vectors. The out-of-plane paths are indicated by black dots.}
\label{sfig8}
\end{figure}

\begin{table}[h]
\centering
\begin{tabular}{c|c|c|c|c|c|c|c|c|c|c|c}
\hline
     & $J^{\pm}_1$ (K) & $J^{\pm\pm}_1$ & $J^{z\pm}_1$ & $J^{\pm}_2$ & $J^{\pm}_3$ & $J^{'\pm}_0$ & $J^{'\pm}_1$ & $J^{'\pm}_{2a}$ & $D$ & scale & damping   \\ \hline
Mean & -1.385     & -1.010       & -3.017       & 0.136       & 0.977      & 0.167        & 0.085         & 0.360          & 25.729 & 0.760    &    0.128  \\ \hline
SD & 0.009       & 0.005          & 0.006        & 0.009       & 0.006       & 0.005        & 0.002        & 0.004           & 0.020 &   0.006    &     0.004  \\ \hline
\end{tabular}\\
\quad\\
\quad\\
\begin{tabular}{c|c|c|c|c|c|c|c}
\hline
     & $J^{zz}_1$ & $J^{zz}_2$ & $J^{zz}_3$ & $J^{'zz}_0$ & $J^{'zz}_1$ & $J^{'zz}_{2a}$ & $E_{\text{SIBS}}$ \\ \hline
Mean & -2.461     & 0.719      & 4.720      & 0.000       & 0.000       & 0.143          & 2.822                   \\ \hline
SD & 0.895      & 0.993      & 0.990       & -           & -           & 0.396          & 0.001                 \\ \hline
\end{tabular}
\caption{The best GSWT fitting parameters of inelastic neutron-scattering data.}
\end{table}

\clearpage

\begin{figure}[ht]
\centering
\includegraphics[width = 0.7\textwidth]{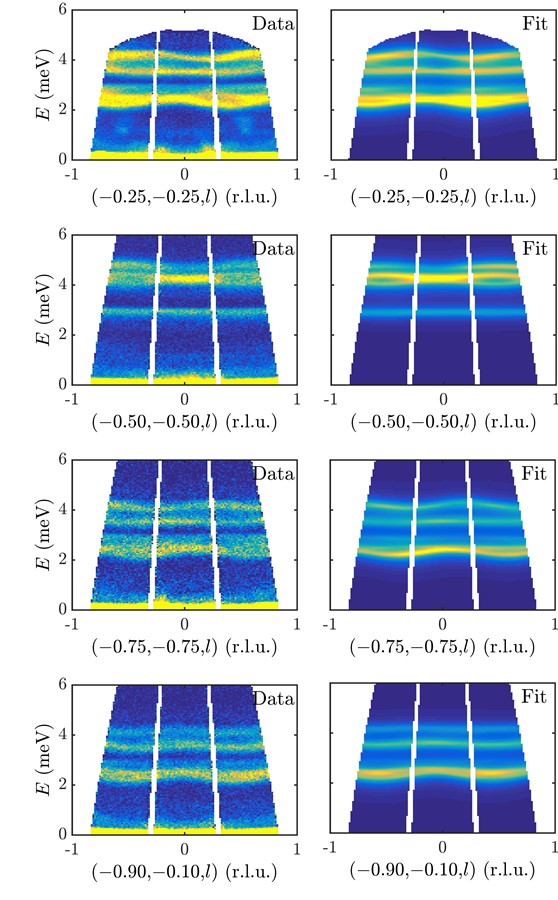}
\caption{The comparison between the data and GSWT fitting for cuts in the out-of-plane direction.}
\label{sfig9}
\end{figure}

\clearpage
\section{Symmetry analysis of exchange Hamiltonian}

The maximally symmetry-allowed exchange interaction on a certain bond can be obtained from SpinW \cite{si-toth2015linear}. It is useful to understand how the symmetry analysis works and double-check the result from SpinW. Here we present two examples to illustrate some of the details. Under a generic symmetry operation $\bU$, spins and atomic positions transform as axial and polar vectors 
\begin{align}
{\bS_{i'} = \text{det}(\bU) \bU\cdot\bS_i\,,\quad \br_{i'}=\bU\cdot\br_i+\textbf{t}.}
\end{align}
The energy of a bond is a scalar, therefore invariant under this operation
\begin{align}
{\bS_{i'}^{T}\cdot \bJ_{i'j'}\cdot\bS_{j'} = \bS_{i}^{T}\cdot \bJ_{ij}\cdot\bS_{j}\,.}
\end{align}
This leads to a transformation of the exchange matrix
\begin{align}
\bJ_{i'j'} = \bU\cdot\bJ_{ij}\cdot\bU^{T}\,.
\end{align}

For the $J_1$ bond between atom $i$ at $(0,0,0)$ and atom $j$ at $(1,0,0)$, the first step in the symmetry analysis is to find the point group element at the center of the bond $(1/2,0,0)$. This information is tabulated on Bilbao crystallographic server under ``Wyckoff Positions". There are three symmetry operations besides the identity. First, let's look at inversion operation $(-x+1,-y,-z)$ which simply switches $(0,0,0)$ and $(1,0,0)$, therefore we have 
\begin{align}
\bJ_{i'j'} =  \bJ_{ji} =  \bU\cdot\bJ_{ij}\cdot\bU^{T} = \bJ_{ij}\,,\quad 
\bU = \textbf{M}\cdot\left(\begin{array}{ccc}
-1& 0&0\\
0& -1&0\\
0& 0&-1
\end{array}\right)\cdot\textbf{M}^{-1}=\left(\begin{array}{ccc}
-1& 0&0\\
0& -1&0\\
0& 0&-1
\end{array}\right)\,.
\end{align}
We can deduce that $\bJ_{ij}$ has to be symmetric, namely, for a bond with inversion 
center the antisymmetric {Dzyaloshinskii-Moriya} interaction is absent. The 
matrix $\textbf{M}$ transforms group elements from the fractional coordinate to the global 
Cartesian coordinate 
\begin{align}
\textbf{M} = \left(
\begin{array}{ccc}
 a & b \cos (\gamma ) & c \cos (\beta ) \\
 0 & b \sin (\gamma ) & c (\cos (\alpha )-\cos (\beta ) \cos (\gamma )) \csc (\gamma ) \\
 0 & 0 & \frac{V \csc (\gamma )}{a b} \\
\end{array}
\right)\,.
\end{align}
The convention of the global Cartesian coordinate is chosen such that the cell vector \textbf{a} is in the positive x-axis direction, the cell vector \textbf{b} in the x-y plane with positive y-axis component, and the cell vector \textbf{c} with positive z-axis component. This choice is consistent with that in \ref{Sec.S4}.
The next symmetry element $(x-y,-y,-z)$ leaves atom $i$ and $j$ unchanged,
\begin{align}
\bJ_{ij} = \bU\cdot\bJ_{ij}\cdot\bU^{T}\,,\quad 
\bU = \textbf{M}\cdot\left(\begin{array}{ccc}
1& -1&0\\
0& -1&0\\
0& 0&-1
\end{array}\right)\cdot\textbf{M}^{-1} =\left(\begin{array}{ccc}
1& 0&0\\
0& -1&0\\
0& 0&-1
\end{array}\right)\,,
\end{align}
where the action $\textbf{M}$ is non-trivial. The exchange matrix is reduced to
\begin{align}\label{eq_J1}
\bJ_{ij} = \left(\begin{array}{ccc}
J_1^{xx}&0 &0\\
0& J_1^{yy}&J_1^{yz}\\
0& J_1^{yz}&J_1^{zz}
\end{array}\right)\,.
\end{align}
The last symmetry operation does not provide further reduction. 
The Hamiltonian in Eq.\ref{eq:NNHam} is obtained by substituting
\begin{align}
J_1^{xx} = 2(J_1^{\pm}+J_1^{\pm\pm})\,,\quad J_1^{yy} = 2(J_1^{\pm}-J_1^{\pm\pm})\,,\quad\text{and}\quad J_1^{yz} = J_1^{z\pm}
\end{align}
and 
\begin{align}
S_i^{x} = \dfrac{1}{2}(S^{+}_{i} + S^{-}_{i})\quad\text{and}\quad S_i^{y} = \dfrac{1}{2i}(S^{+}_{i} - S^{-}_{i}) 
\end{align}
in
\begin{align}
\bS_i\cdot\bJ_{ij}\cdot \bS_j = J_{1}^{xx} S_i^{x}S_{j}^{x}+ J_{1}^{yy} S_i^{y}S_{j}^{y}+ J_{1}^{yz} (S_i^{z}S_{j}^{y}+S_i^{y}S_{j}^{z})+ J_{1}^{zz} S_i^{z}S_{j}^{z}\,,
\end{align}
which gives
\begin{align}
J_{1}^{\pm}(S_{i}^{+} S_{j}^{-} + S_{i}^{-} S_{j}^{+}) + J_{1}^{\pm\pm} (S_{i}^{+} S_{j}^{+} + S_{i}^{-} S_{j}^{-}) + J_{1}^{zz} S_{i}^{z} S_{j}^{z} 
 -\dfrac{i J_z^{\pm}}{2}  (( S_{i}^{+} - S_{i}^{-} )S_{j}^{z} +S_i^z( S_j^+- S_j^- ))\,.
\end{align}
To obtain the exchange matrix for the other $J_1$ bonds, we just need to act corresponding symmetry operation on Eq.~\ref{eq_J1}, then use the same substitution above. For a general rotation of angle $\theta$ along z-axis, the result reads
\begin{align}
&J_{1}^{\pm}(S_{i}^{+} S_{j}^{-} + S_{i}^{-} S_{j}^{+}) + J_{1}^{\pm\pm} (e^{-2i\theta} S_{i}^{+} S_{j}^{+} + e^{2i\theta} S_{i}^{-} S_{j}^{-} ) + J_{1}^{zz} S_{i}^{z} S_{j}^{z} 
\notag\\
& -\dfrac{i J_z^{\pm}}{2}  ((e^{-i\theta}  S_{i}^{+} - e^{i\theta} S_{i}^{-} )S_{j}^{z} + S_i^z(e^{-i\theta}  S_j^+-e^{i\theta}  S_j^- ))\,.
\end{align}
As a second example, it is instructive to look at a $J^{'}_{1}$ bond between atom $i$ at $(0,0,0)$ and atom $j$ at $(0,0,1)$. Only the symmetry operation $(-x+1,-y+1,-z)$ maps this bond onto itself up to a lattice translation, leading to $\bJ_{ij} = \bJ_{ij}^{T}$. The other two operations map it to other $J^{'}_{1}$ bonds, therefore does not place further restriction on the exchange matrix, leaving six symmetry-allowed parameters for this bond. This result agrees with calculation from SpinW.

\clearpage
\subsection*{Supplementary Information References}
{\def\section*#1{}
\renewcommand\refname{}

}


\begin{thebibliography}{10}

\bibitem{savary2016quantum}
L.~Savary, L.~Balents, {\it Reports on Progress in Physics\/} {\bf 80}, 016502
  (2016).

\bibitem{Nakatsuji2005}
S. Nakatsuji, Y. Nambu, H. Tonomura, O. Sakai, S. Jonas, C. L. Broholm,  H. Tsunetsugu, Y. Qiu, Y. Maeno,
  {\it Science\/} {\bf 309}, 1697 (2005).

\bibitem{Broholmeaay0668}
C.~Broholm, R.~J. Cava, S.~A. Kivelson, D.~G. Nocera, M.~R. Norman, T.~Senthil,
  {\it Science\/} {\bf 367} (2020).

\bibitem{bertrand1974susceptibilite}
Y.~Bertrand, A.~Fert, J.~Gelard, {\it Journal de Physique\/} {\bf 35}, 385
  (1974).

\bibitem{batista2004algebraic}
C.~D. Batista, G.~Ortiz, {\it Advances in Physics\/} {\bf 53}, 1 (2004).

\bibitem{Takagi2019}
  H.~Takagi, T.~Takayama, G.~Jackeli, G.~Khaliullin, S.~E.~Nagler, {\it Nature Review Physics\/} {\bf 1}, 264--280 (2019).

\bibitem{suzuki2018first}
M.-T. Suzuki, H.~Ikeda, P.~M. Oppeneer, {\it Journal of the Physical Society of
  Japan\/} {\bf 87}, 041008 (2018).

\bibitem{kuramoto2009multipole}
Y.~Kuramoto, H.~Kusunose, A.~Kiss, {\it Journal of the Physical Society of
  Japan\/} {\bf 78}, 072001 (2009).

\bibitem{santini2009multipolar}
P.~Santini, S.~Carretta, G.~Amoretti, R.~Caciuffo, N.~Magnani, G.~H. Lander,
  {\it Reviews of Modern Physics\/} {\bf 81}, 807 (2009).

\bibitem{matsumoto2007longitudinal}
M.~Matsumoto, M.~Koga, {\it Journal of the Physical Society of Japan\/} {\bf
  76}, 073709 (2007).

\bibitem{romhanyi2012multiboson}
J.~Romh{\'a}nyi, K.~Penc, {\it Physical Review B\/} {\bf 86}, 174428 (2012).

\bibitem{fert1978excitation}
A.~Fert, D.~Bertrand, J.~Leotin, J.~Ousset, J.~Magari{\~n}o, J.~Tuchendler,
  {\it Solid State Communications\/} {\bf 26}, 693 (1978).

\bibitem{petitgrand1980magnetic}
D.~Petitgrand, A.~Brun, P.~Meyer, {\it Journal of Magnetism and Magnetic
  Materials\/} {\bf 15}, 381 (1980).

\bibitem{silberglitt1970effect}
R.~Silberglitt, J.~B. Torrance~Jr, {\it Physical Review B\/} {\bf 2}, 772
  (1970).

\bibitem{oguchi1971theory}
T.~Oguchi, {\it Journal of the Physical Society of Japan\/} {\bf 31}, 394
  (1971).

\bibitem{petitgrand1979neutron}
D.~Petitgrand, B.~Hennion, C.~Escribe, {\it Journal of Magnetism and Magnetic
  Materials\/} {\bf 14}, 275 (1979).

\bibitem{katsumata2000single}
K.~Katsumata, H.~Yamaguchi, M.~Hagiwara, M.~Tokunaga, H.-J. Mikeska, P.~Goy,
  M.~Gross, {\it Physical Review B\/} {\bf 61}, 11632 (2000).

\bibitem{balucani1985hybrid}
U.~Balucani, A.~Stasch, {\it Physical Review B\/} {\bf 32}, 182 (1985).

\bibitem{fujita1966mossbauer}
T.~Fujita, A.~Ito, K.~{\^O}no, {\it Journal of the Physical Society of Japan\/}
  {\bf 21}, 1734 (1966).

\bibitem{gelard1974magnetic}
J.~Gelard, A.~Fert, P.~Meriel, Y.~Allain, {\it Solid State Communications\/}
  {\bf 14}, 187 (1974).

\bibitem{wiedenmann1988neutron}
A.~Wiedenmann, L.~Regnault, P.~Burlet, J.~Rossat-Mignod, O.~Kound{\'e},
  D.~Billerey, {\it Journal of magnetism and magnetic materials\/} {\bf 74}, 7
  (1988).

\bibitem{tanaka1975ground}
Y.~Tanaka, N.~Ury{\^u}, {\it Journal of the Physical Society of Japan\/} {\bf
  39}, 825 (1975).

\bibitem{lockwood1994raman}
D.~Lockwood, G.~Mischler, A.~Zwick, {\it Journal of Physics: Condensed
  Matter\/} {\bf 6}, 6515 (1994).

\bibitem{conlon2010absent}
P.~Conlon, J.~Chalker, {\it Physical Review B\/} {\bf 81}, 224413 (2010).

\bibitem{plumb2019continuum}
K.~Plumb, H.~J. Changlani, A.~Scheie, S.~Zhang, J.~Krizan, J.~Rodriguez-Rivera,
  Y.~Qiu, B.~Winn, R.~J. Cava, C.~Broholm, {\it Nature Physics\/} {\bf 15}, 54
  (2019).

\bibitem{bai2019magnetic}
X.~Bai, J.~Paddison, E.~Kapit, S.~Koohpayeh, J.-J. Wen, S.~Dutton, A.~Savici,
  A.~Kolesnikov, G.~Granroth, C.~Broholm, {\it et~al.\/}, {\it Physical Review
  Letters\/} {\bf 122}, 097201 (2019).

\bibitem{Wu2012}
X.~Wu, Y.~Cai, Q.~Xie, H.~Weng, H.~Fan, J.~Hu, {\it Physical Review B\/} {\bf
  86}, 1 (2012).

\bibitem{muniz2014generalized}
R.~A. Muniz, Y.~Kato, C.~D. Batista, {\it Progress of Theoretical and
  Experimental Physics\/} {\bf 2014} (2014).

\bibitem{paddison2020scattering}
J.~A. Paddison, {\it arXiv preprint arXiv:2002.12894\/}  (2020).

\bibitem{li2015rare}
Y.~Li, G.~Chen, W.~Tong, L.~Pi, J.~Liu, Z.~Yang, X.~Wang, Q.~Zhang, {\it
  Physical review letters\/} {\bf 115}, 167203 (2015).

\bibitem{maksimov2019anisotropic}
P.~Maksimov, Z.~Zhu, S.~R. White, A.~Chernyshev, {\it Physical Review X\/} {\bf
  9}, 021017 (2019).

\bibitem{paddison2017continuous}
J.~A. Paddison, M.~Daum, Z.~Dun, G.~Ehlers, Y.~Liu, M.~B. Stone, H.~Zhou,
  M.~Mourigal, {\it Nature Physics\/} {\bf 13}, 117 (2017).

\bibitem{penc2012spin}
K.~Penc, J.~Romh{\'a}nyi, T.~R{\~o}{\~o}m, U.~Nagel, {\'A}.~Antal,
  T.~Feh{\'e}r, A.~J{\'a}nossy, H.~Engelkamp, H.~Murakawa, Y.~Tokura, {\it
  et~al.\/}, {\it Physical review letters\/} {\bf 108}, 257203 (2012).

\bibitem{akaki2017direct}
M.~Akaki, D.~Yoshizawa, A.~Okutani, T.~Kida, J.~Romh{\'a}nyi, K.~Penc,
  M.~Hagiwara, {\it Physical Review B\/} {\bf 96}, 214406 (2017).

\bibitem{zvyagin2008observation}
S.~Zvyagin, C.~Batista, J.~Krzystek, V.~Zapf, M.~Jaime, A.~Paduan-Filho,
  J.~Wosnitza, {\it Physica B: Condensed Matter\/} {\bf 403}, 1497 (2008).

\bibitem{yoshizawa1980neutron}
H.~Yoshizawa, W.~Kozukue, K.~Hirakawa, {\it Journal of the Physical Society of
  Japan\/} {\bf 49}, 144 (1980).

\bibitem{hayashida2019novel}
S.~Hayashida, M.~Matsumoto, M.~Hagihala, N.~Kurita, H.~Tanaka, S.~Itoh,
  T.~Hong, M.~Soda, Y.~Uwatoko, T.~Masuda, {\it arXiv preprint
  arXiv:1908.08403\/}  (2019).

\bibitem{fert1973phase}
A.~Fert, J.~Gelard, P.~Carrara, {\it Solid State Communications\/} {\bf 13},
  1219 (1973).

\bibitem{wiedenmann1989magnetic}
A.~Wiedenmann, L.~Regnault, P.~Burlet, J.~Rossat-Mignod, O.~Kounde,
  D.~Billerey, {\it Physica B: Condensed Matter\/} {\bf 156}, 305 (1989).

\bibitem{dally2020high}
R.~L. Dally, A.~J. Heng, A.~Keselman, M.~M. Bordelon, M.~B. Stone, L.~Balents,
  S.~D. Wilson, {\it arXiv preprint arXiv:2001.07300\/}  (2020).

\end{thebibliography}

\begin{thebibliography}{1} 

\bibitem{method-coleman1993optimization}
C.~Coleman, E.~Yamada, {\it Journal of Crystal Growth\/} {\bf 132}, 129 (1993).

\bibitem{method-petitgrand1976far}
D.~Petitgrand, P.~Meyer, {\it Journal de Physique\/} {\bf 37}, 1417 (1976).

\bibitem{method-friedt1976electronic}
J.~Friedt, J.~Sanchez, G.~Shenoy, {\it The Journal of Chemical Physics\/} {\bf
  65}, 5093 (1976).

\bibitem{method-trooster1978spin}
J.~Trooster, W.~de~Valk, {\it Hyperfine Interactions\/} {\bf 4}, 457 (1978).

\bibitem{method-katsumata2000observation}
K.~Katsumata, M.~Hagiwara, M.~Tokunaga, H.~Yamaguchi, {\it Journal of Applied
  Physics\/} {\bf 87}, 5085 (2000).

\bibitem{method-rodriguez1993recent}
J.~Rodr{\'\i}guez-Carvajal, {\it Physica B: Condensed Matter\/} {\bf 192}, 55
  (1993).

\bibitem{method-ye2018implementation}
F.~Ye, Y.~Liu, R.~Whitfield, R.~Osborn, S.~Rosenkranz, {\it Journal of Applied
  Crystallography\/} {\bf 51}, 315 (2018).

\bibitem{method-granroth2010sequoia}
G.~Granroth, A.~Kolesnikov, T.~Sherline, J.~Clancy, K.~Ross, J.~Ruff,
  B.~Gaulin, S.~Nagler, {\it Journal of Physics: Conference Series\/} (IOP
  Publishing, 2010), vol. 251, p. 012058.

\bibitem{method-stone2014comparison}
M.~B. Stone, J.~L. Niedziela, D.~L. Abernathy, L.~DeBeer-Schmitt, G.~Ehlers,
  O.~Garlea, G.~Granroth, M.~Graves-Brook, A.~I. Kolesnikov, A.~Podlesnyak,
  {\it et~al.\/}, {\it Review of Scientific Instruments\/} {\bf 85}, 045113
  (2014).

\bibitem{method-arnold2014mantid}
O.~Arnold, J.-C. Bilheux, J.~Borreguero, A.~Buts, S.~I. Campbell, L.~Chapon,
  M.~Doucet, N.~Draper, R.~F. Leal, M.~Gigg, {\it et~al.\/}, {\it Nuclear
  Instruments and Methods in Physics Research Section A: Accelerators,
  Spectrometers, Detectors and Associated Equipment\/} {\bf 764}, 156 (2014).

\bibitem{method-ewings2016horace}
R.~Ewings, A.~Buts, M.~Le, J.~van Duijn, I.~Bustinduy, T.~Perring, {\it Nuclear
  Instruments and Methods in Physics Research Section A: Accelerators,
  Spectrometers, Detectors and Associated Equipment\/} {\bf 834}, 132 (2016).

\end{thebibliography}

\begin{thebibliography}{1}

\bibitem{si-aroyo2011crystallography}
M.~I. Aroyo, J.~Perez-Mato, D.~Orobengoa, E.~Tasci, G.~De~La~Flor, A.~Kirov,
  {\it Bulg. Chem. Commun\/} {\bf 43}, 183 (2011).

\bibitem{si-ewings2016horace}
R.~Ewings, A.~Buts, M.~Le, J.~van Duijn, I.~Bustinduy, T.~Perring, {\it Nuclear
  Instruments and Methods in Physics Research Section A: Accelerators,
  Spectrometers, Detectors and Associated Equipment\/} {\bf 834}, 132 (2016).

\bibitem{si-wiedenmann1988neutron}
A.~Wiedenmann, L.~Regnault, P.~Burlet, J.~Rossat-Mignod, O.~Kound{\'e},
  D.~Billerey, {\it Journal of Magnetism and Magnetic Materials\/} {\bf 74}, 7
  (1988).

\bibitem{si-conlon2010absent}
P.~Conlon, J.~Chalker, {\it Physical Review B\/} {\bf 81}, 224413 (2010).

\bibitem{si-bai2019magnetic}
X.~Bai, J.~Paddison, E.~Kapit, S.~Koohpayeh, J.-J. Wen, S.~Dutton, A.~Savici,
  A.~Kolesnikov, G.~Granroth, C.~Broholm, {\it et~al.\/}, {\it Physical Review
  Letters\/} {\bf 122}, 097201 (2019).

\bibitem{si-plumb2019continuum}
K.~Plumb, H.~J. Changlani, A.~Scheie, S.~Zhang, J.~Krizan, J.~Rodriguez-Rivera,
  Y.~Qiu, B.~Winn, R.~J. Cava, C.~Broholm, {\it Nature Physics\/} {\bf 15}, 54
  (2019).

\bibitem{si-nlopt}
S.~G. Johnson, The NLopt nonlinear-optimization package (2008--2019).

\bibitem{si-nelder1965simplex}
J.~A. Nelder, R.~Mead, {\it The Computer Journal\/} {\bf 7}, 308 (1965).

\bibitem{si-toth2015linear}
S.~Toth, B.~Lake, {\it Journal of Physics: Condensed Matter\/} {\bf 27}, 166002
  (2015).

\end{thebibliography}
\end{document}